\documentclass[traditabstract]{aa}
\usepackage{graphicx}
\usepackage{txfonts}
\usepackage{longtable}
\begin{document}
   \title{The GALEX Ultraviolet Virgo Cluster Survey (GUViCS). IV: The role of the cluster environment on galaxy evolution}
   \subtitle{}
  \author{A. Boselli\inst{1}
          ,
	  E. Voyer\inst{1}
	  ,
	  S. Boissier\inst{1,2}
	  ,
	  O. Cucciati\inst{3,2}
	  ,
	  G. Consolandi\inst{4}
	  ,
	  L. Cortese\inst{5}
	  ,
	  M. Fumagalli\inst{6,9}
	  ,
	  G. Gavazzi\inst{4}
	  ,
	  S. Heinis\inst{7}
          ,
 	  Y. Roehlly\inst{1}
	  ,
	  E. Toloba\inst{8,9}
       }

\institute{	
		Aix Marseille Universit\'e, CNRS, LAM (Laboratoire d'Astrophysique de Marseille), UMR 7326, F-13388, Marseille, France
             \email{Alessandro.Boselli@lam.fr, Samuel.Boissier@lam.fr, elysse.voyer@gmail.com, yannick.roehlly@lam.fr}
         \and  
	 	I.N.A.F., Osservatorio Astronomico di Bologna, via Ranzani 1, I-40127 Bologna, Italy
	 \and	
	 	Universita' di Bologna, Dipartimento di Fisica e Astronomia, viale Berti Pichat 6/2, 40127, Bologna, Italy
		\email{olga.cucciati@oabo.inaf.it}
	 \and
	        Universita' di Milano-Bicocca, piazza della Scienza 3, 20100, Milano, Italy
             \email{guido.consolandi@mib.infn.it, giuseppe.gavazzi@mib.infn.it}
	 \and
	 	Centre for Astrophysics \& Supercomputing, Swinburne University of Technology, Mail H30, PO Box 218, Hawthorn, VIC 3122, Australia
             \email{lcortese@swin.edu.au}
	 \and
	        Institute for Computational Cosmology, Department of Physics, Durham University, South Road, Durham DH1 3LE, UK
	     \email{michele.fumagalli@durham.ac.uk}
	 \and
	        Department of Astronomy, University of Maryland, College Park, MD 20742-2421
	     \email{sheinis@astro.umd.edu}
	 \and
	        UCO/Lick Observatory, University of California, Santa Cruz, 1156 High Street, Santa Cruz, CA 95064, USA
	     \email{toloba@ucolick.org}
	 \and{Carnegie Observatories, 813 Santa Barbara Street, Pasadena, CA 91101, USA}
}

\authorrunning{Boselli et al.}
\titlerunning{GUViCS: the role of the cluster environment on galaxy evolution}

   \date{}

 
  \abstract  
{We study the role of the environment on galaxy evolution using a sample of 868 galaxies in the Virgo cluster and in its surrounding regions selected from the 
\textit{GALEX} Ultraviolet Virgo Cluster Survey (GUViCS) with the purpose of understanding the origin of the red sequence in dense environments.
The sample spans a wide range in morphological types (from dwarf ellipticals to Im and BCD) and stellar masses 
(10$^7$ $\lesssim$ $M_{star}$ $\lesssim$ 10$^{11.5}$ M$_{\odot}$). We collected multifrequency data covering the whole electromagnetic spectrum for most of the galaxies, including UV, optical, 
mid- and far-infrared imaging data as well as optical and HI spectroscopic data. We first identify the different dynamical substructures composing the Virgo cluster and we calculate the 
local density of galaxies using different methods. We then study the distribution of galaxies belonging to the red sequence,
the green valley, and the blue cloud within the different cluster substructures or as a function of galaxy density. Our analysis indicates that all the most massive galaxies 
($M_{star}$ $\gtrsim$ 10$^{11}$ M$_{\odot}$) are slow rotators and are the dominant galaxies of the different cluster substructures generally associated with a diffuse X-ray emission.
They are probably the result of major merging events that occurred at early epochs, as also indicated by their very old stellar populations.
Slow rotators of lower stellar mass (10$^{8.5}$ $\lesssim$ $M_{star}$ $\lesssim$ 10$^{11}$ M$_{\odot}$) are also preferentially located within the different high-density substructures of the cluster.
Their position in the velocity space indicates that they are virialised within the cluster, thus Virgo members since its formation.
They have been shaped by gravitational perturbations occurring within the infalling groups that later formed the cluster (pre-processing).
On the contrary, low-mass star-forming systems are extremely rare in the inner regions of the Virgo cluster A, where
the density of the intergalactic medium is at its maximum. Our ram pressure stripping models consistently indicate that these star-forming systems
can be rapidly deprived of their interstellar medium during their interaction with the intergalactic medium. The lack of gas quenches their star formation activity 
transforming them into quiescent dwarf ellipticals. This mild transformation does not perturb the kinematic properties of these galaxies which still have rotation curves typical of star-forming systems.

 }
   {}
   {}
   {}
   {}
   {}


   \maketitle
%

\section{Introduction}

Multifrequency observations of nearby and high-redshift galaxies consistently indicate that mass is the principal driver of galaxy evolution (down-sizing effect; 
Cowie et al. 1996; Gavazzi et al. 1996; Boselli et al. 2001; Fontanot et al. 2009). Massive galaxies have formed most of their stars at early epochs, while 
dwarf systems are still active at a rate comparable to their mean star formation rate during all their life. 
Observations, however, clearly indicate that mass is not the only parameter driving galaxy evolution. There is indeed strong observational 
evidence suggesting that the environment in which galaxies reside might be another key parameter. Since the seminal work of Dressler we know that 
galaxies in high-density environments are preferentially ellipticals and lenticulars, the former dominating the core of rich clusters (morphology segregation effect; 
Dressler et al. 1980; 1997; Whitmore et al. 1993). There is also evidence that field galaxies, mainly late-type gas-rich systems, are falling into high-density 
regions (Colless \& Dunn 1996; Rines et al. 2003). What is the fate of these freshly infalling systems in high density environments? 

Clusters of galaxies are high-density environments characterised by a deep potential well trapping a hot and dense intergalactic medium emitting in the X-rays.
The gravitational interactions of galaxies with other cluster members or with the potential of the cluster as a whole, as well as their interactions
with the diffuse intergalactic medium, can easily remove their interstellar medium quenching their star formation activity because of the lack of fresh fuel. 
Spiral galaxies can thus be transformed into quiescent systems (e.g. Boselli \& Gavazzi 2006).
While the main lines of this evolutionary picture through cosmic time are rather well understood (Elbaz et al. 2007; Bundy et al. 2010), we still do not know the 
exact contribution of each perturbing process since the formation of galaxies to the present epoch. Indeed we still do not know which are the physical 
processes that gave birth to the different galaxy populations inhabiting rich clusters: massive and dwarf ellipticals, lenticulars and anemic spirals. We also do 
not know how the relative weight of the different processes that shape galaxy evolution changed since their formation.
This is indeed expected given that the physical conditions characterising high-density regions (total mass, velocity dispersion, density of the intergalactic medium) 
significantly changed with cosmic time.

The Virgo cluster is the highest density region close to the Milky Way. It is a cluster still in formation composed of different substructures similar to those
expected in high redshift clusters. These substructures are quite different from one another, since they span 
a wide range in velocity dispersion, galaxy composition and properties of the intergalactic medium. Virgo is thus an ideal laboratory for 
studying and comparing the effects induced by different kinds of perturbations.
Furthermore, thanks to its proximity ($\sim$ 16.5-17 Mpc; Gavazzi et al. 1999; Mei et al. 2007), observations of Virgo dwarf galaxies are possible at almost 
any frequency. This is important since these are the most fragile objects easily perturbed in any kind of physical process. 
At the same time, the angular dimension of galaxies is sufficiently large to allow 
the detailed comparison of their radial properties with the prediction of different models of galaxy evolution. This is of paramount importance
for the identification of the ongoing perturbing process (e.g. Boselli et al. 2006).

For all these reasons the Virgo cluster has always been one of the preferred targets in environmental studies. After the seminal work of Binggeli, Sandage, and Tammann,  
(e.g. Binggeli et al. 1985), who mapped the whole cluster region in one optical band using photographic plates, however, blind surveys of the Virgo cluster
were not possible up to the last years because of its large extension on the sky (more than 100 deg.$^2$). Dedicated studies were thus focused on selected
samples of galaxies for which multifrequency data were becoming available. It is only in the recent years that the advent of large panoramic detectors
allowed the full mapping of the Virgo cluster in several photometric and spectroscopic bands. 
Multifrequency observations are crucial since they provide information on the different
components of galaxies, including both the young and old stellar populations, the different constituents of the interstellar medium (ISM; atomic and molecular gas,
dust, metals), magnetic fields etc. They are thus a unique tool to study how the matter cycle in galaxies is perturbed in high-density environments (e.g.
Boselli 2011). Several blind surveys of the Virgo cluster have been recently completed in the optical bands (NGVS; Ferrarese et al. 2012), in the mid-
(\textit{WISE}; Wright et al. 2010) and far-infrared (HeViCS; Davies et al. 2010; 2012), and in the 21 cm HI line (ALFALFA; Giovanelli et al. 2005). 
The \textit{GALEX} Ultraviolet Virgo Cluster Survey (GUViCS; Boselli et al. 2011), a deep blind survey in two UV photometric bands of $\sim$ 300 deg.$^2$ centered on M87,
has been recently completed. UV data are sensitive to the emission of the youngest 
stars in star-forming systems (e.g. Kennicutt 1998a; Boselli et al. 2009) and to that of the most evolved stars in old, early-type galaxies (O'Connell 1999;
Boselli et al. 2005). They are thus crucial for reconstructing the recent and past star formation history of perturbed and unperturbed objects in the nearby
universe.

It is thus time to revisit the seminal work of Sandage and collaborators and extend the study of the Virgo cluster taking benefit
of the unique set of multifrequency photometric and spectroscopic data now available to the community.
In this paper we analyse the statistical properties of a large sample of more than eight hundred galaxies located in the Virgo cluster and 
in its surroundings.
The effectiveness of using a statistical analysis based on multifrequency data in the study of the role of the environment on galaxy evolution has been recently
shown in the works of Gavazzi et al. (2013a,b), which combined HI, H$\alpha$ and optical data in the Local supercluster, including Virgo, and in the Coma
supercluster region. More recently, Cybulski et al. (2014) combined near- and mid-infrared data from the \textit{WISE} survey (Wright et al. 2010) with UV data from
\textit{GALEX} to study the star formation history of galaxies in the Coma supercluster region. Here we combine the new set of UV data 
from \textit{GALEX} recently published in Voyer et al. (2014) with SDSS optical data, mid- and far-infrared data from 
\textit{WISE} and \textit{Herschel}, and HI data from ALFALFA to have a complete picture of galaxy evolution
within the Virgo cluster region. Additionally, we add a few high resolution spectroscopic data that are extremely useful in quantitatively estimating the kinematic
properties of a representative subsample of massive and intermediate mass early-type systems. The results of this analysis are compared to the predictions
of multizone chemo-spectrophotometric models of galaxy evolution presented in Boselli et al. (2006, 2008a), specially tailored to take into account two different processes induced by the cluster 
environment on galaxies: ram pressure stripping (Gunn \& Gott 1972) and starvation (Larson et al. 1980). The study of the origin of the red sequence through 
the transformation of late-type galaxies in high-density environments has been the topic of several recent papers (e.g. Boselli et al. 2008a, Hughes \& Cortese 2009, Cortese \& Hughes 2009,
Gavazzi et al. 2010, 2013a, 2013b). The novelty this work is at the same time that of using the largest sample with a complete set of 
multifrequency data spanning the whole electromagnetic spectrum, extending previous analyses down to dwarf galaxies of stellar mass $M_{star}$ $\simeq$ 10$^7$ M$_{\odot}$,
and taking benefit of the proximity of Virgo to resolve galaxies within its cluster substructures where different physical processes are dominant.

The paper is structured as follows: in Section 2 we present the sample, in Section 3 the multifrequency set of data used in the analysis, while in Section 4
we briefly describe our multizone chemo-spectrophotometric models of galaxy evolution. In Section 5 we use the UV-to-optical colour magnitude relation to
characterise the different galaxy populations, while in Section 6 we study the distribution of the different types of galaxies within the various substructures
of the cluster. The analysis is presented in Section 7, while a detailed discussion of the results is given in Section 8. In Appendix A we present the new
set of \textit{WISE} data at 22 $\mu$m necessary to correct for dust attenuation the UV emission of the target galaxies, while in Appendix B
we study how the use of standard recipes for determining the total stellar mass of galaxies might induce systematic effects in perturbed objects.

\section{The sample}

The sample analysed in this work has been extracted from the Extended Source Catalogue of Voyer et al. (2014)
and it is composed of all galaxies detected by \textit{GALEX} in the NUV band in the Virgo cluster region and 
its surroundings (12h $\leq$ $R.A.$ $\leq$ 13h; 0$^o$ $\leq$ $dec$ $\leq$ 20$^o$) with a recessional velocity lower than 3500 km s$^{-1}$. 
The Extended Source Catalogue of Voyer et al. (2014) is composed of all galaxies listed in the Virgo Cluster Catalogue (VCC, Binggeli et al. 1985), 
in the CGCG (Zwicky et al. 1961-1968), or in other main catalogues (NGC, UGC, IC, DDO, KUG, FGC, MRK, LSBC, AGC ...) included in NED.  
The adopted threshold in redshift guarantees the inclusion of galaxies in the infalling regions and in the different substructures of Virgo
mainly located at slightly higher distance than the main body of the cluster associated to M87, generally called cluster A (Gavazzi et al. 1999).

As defined, the analysed sample is optical and UV selected and can thus suffer from incompleteness in the two bands. In the optical band the catalogues used to extract
the target galaxies are not complete at the same depth over the studied region.
The VCC, which covers the largest portion of the sky analysed in this work, is complete to the photographic magnitude $m_{pg}$ $\simeq$ 18. Down to this magnitude limit
the VCC is also almost complete in redshift (88\%).
Boselli et al. (2011) have shown that at the typical depth of the deep GUViCS observations (Medium Imaging Survey), 92 \% ~ of the galaxies detected by \textit{GALEX}
with a NUV magnitude $\leq$ 21 mag have an optical counterpart in the VCC. At this depth, the GUViCS observations cover $\sim$ 65 \% ~ of the studied region
and encompass the full VCC (Fig. \ref{angdistmass1}). This region is slightly more extended than the virial radius of cluster A and B (the substructure associated to M49). 
In this region the NUV catalogue is complete down to $\sim$ 21.5 AB mag (Voyer et al. 2014).  
The same completeness both in the optical and UV bands is unfortunately not reached in the periphery of the cluster. The CGCG is complete to $m_{pg}$ $\simeq$ 15.7,
while the other catalogues used to define the sample have been constructed using different selection criteria. In the same region, the sky coverage of \textit{GALEX} 
at the depth of the MIS is also more sporadic. At the depth of the All sky Imaging Survey the sample is complete only down to $\sim$ 20 AB mag (see Table \ref{data} and Fig. \ref{angdistmass1}).

The final sample used in this work includes 868 objects down to the NUV limit of $\sim$ 22 AB mag.
The UV selection favors the detection of faint star-forming galaxies down to stellar masses $M_{star}$ $\simeq$ 10$^{6.5}$ M$_{\odot}$ (see sect. 3.2). 
Early-type galaxies, because of their quiescent nature,
have redder colours than star-forming systems and are detected only to $M_{star}$ $\simeq$ 10$^{7}$ M$_{\odot}$. The survey is complete to these stellar mass limits only
at the depth of the MIS, thus up to $\sim$ 1 virial radius of cluster A and B, and within all the other cluster substructures (see section 6.1). 
However, in the periphery of the cluster, where the NUV data comes principally from the AIS, the sample is complete only to $M_{star}$ $\simeq$ 10$^8$-10$^{8.5}$ M$_{\odot}$ 
in quiescent objects with red colours.

\begin{table}
\caption{Completeness in the NUV band over the Virgo cluster region (12h $\leq$ $R.A.$ $\leq$ 13h; 0$^o$ $\leq$ $dec$ $\leq$ 20$^o$).}
\label{data}
{
\[
\begin{tabular}{cccc}
\hline
\noalign{\smallskip}
\hline
Sample	& Depth			& Coverage	& Completeness \\
\hline
AIS	& $\sim$ 200 sec	&  94\%		& 20 mag \\
MIS	& $\gtrsim$ 800 sec	&  65\%		& 21.5 mag \\
\noalign{\smallskip}
\hline
\end{tabular}
\]
}
\end{table}

\section{The data}


\subsection{The multifrequency data}

The Virgo cluster region has been the target of the \textit{GALEX} Virgo Cluster Survey (GUViCS; Boselli et al. 2011).
Because of the adopted selection criteria (see sect. 2), all galaxies have \textit{GALEX} data in the NUV-band ($\rm \lambda_{eff}=2316\AA, \Delta \lambda=1060\AA$), 
while only 531 (62\%) in the FUV-band ($\rm\lambda_{eff}=1539\AA, \Delta \lambda=442\AA$) down to 22 AB mag. 
The UV data have been taken 
from the GUViCS catalogue recently published in Voyer et al. (2014). Being extended sources, the UV flux of galaxies at the distance of Virgo has been extracted using ad
hoc procedures defined to encompass their whole emission. The UV fluxes analysed in this work can thus be considered as total entities.\\

UV data are combined with optical data to constrain the properties of the stellar emission within galaxies. The optical data have been taken in 
the SDSS photometric bands ($u, g, r, i, z$) from the SDSS (Abazajian et al. 2009).
As for the UV bands, to avoid the use of the datasets extracted from the standard pipelines, which are known to suffer important shredding given the extended nature of these nearby galaxies, 
optical magnitudes have been taken from imaging photometry for extended sources specially tuned to measure the total emission of these targets. These data come, in order of preference,
from the SDSS imaging of the \textit{Herschel} Reference Sample (Boselli et al. 2010) recently published by Cortese et al. (2012a), which includes the brightest 172 objects, 
from the compilation of Consolandi et al. (in preparation) determined by fitting composite radial light profiles for galaxies with stellar masses $M_{star}$ $\gtrsim$ 10$^{9.5}$ M$_{\odot}$,
from the set of data determined using a similar procedure by Grossetti (2010) and Galardo (2010), from aperture photometry (Gavazzi et al. 2013a), or from the SDSS standard pipeline
for the remaining faintest objects. Photometric optical data are available for all the galaxies of the sample.
The SDSS also provides nuclear spectra (in a circular aperture of 3 arcsec) for 575 galaxies of the sample. These nuclear spectra are used to identify post-starburst galaxies (PSB or k+a)
using the criterion described in Poggianti et al. (2004) and Dressler et al. (1999) (see however Quintero et al. 2004), i.e.
galaxies with a Balmer absorption line with an equivalent width E.W.H$\delta$ $>$ 3 ~\AA ~for a signal-to-noise larger than 5. 
They are also used to identify galaxies with a nuclear star formation as those objects with a Balmer H$\alpha$ emission line E.W.H$\alpha$ $>$ 3 ~\AA ~(with a signal-to-noise larger than 5).
In bright galaxies, which are not fully sampled in the spectroscopic SDSS survey, this condition might include active galactic nuclei.
For these reasons this nuclear classification will be used in the following analysis only for objects of low and intermediate stellar mass ($M_{star}$ $\leq$ 10$^{9.5}$ M$_{\odot}$).\\

The determination of the dust attenuation in the different photometric bands, in particular in the UV ones, requires the use of infrared data (see below). To gather this information
we extract infrared data in the 22 $\mu$m band from the \textit{WISE} survey (Wright et al. 2010). To avoid any possible 
systematic effect due to the extended nature of the target galaxies, we do not use published catalogues, optimised for point-like sources, but rather extract fluxes from the images using
procedures similar to those adopted in the other bands, as described in Appendix A. \textit{WISE} detections at 22 $\mu$m are available for 407 objects (47\%).  

The UV catalogue analysed in this work has also been cross-matched with the ALFALFA HI survey (Giovanelli et al. 2005) recently published in Haynes et al. (2011). 
Given the limited sensitivity of this survey ($rms$ $\sim$ 2.3 mJy at 5 km s$^{-1}$ spectral resolution), 
which allows the detection of galaxies with $M(HI)$ $\simeq$ 10$^{7.5}$ M$_{\odot}$ at the typical distance of Virgo, we also
cross-matched the UV catalogue with deep HI data collected in the GOLDMine database (Gavazzi et al. 2003), available mainly for late-type galaxies, where the 
sensitivity is up to a factor of $\sim$ 5 higher. HI masses are available for 354 mainly late-type galaxies. Including upper limits, HI data are available for 742 objects.

\begin{figure*}
   \centering
   \includegraphics[width=17cm]{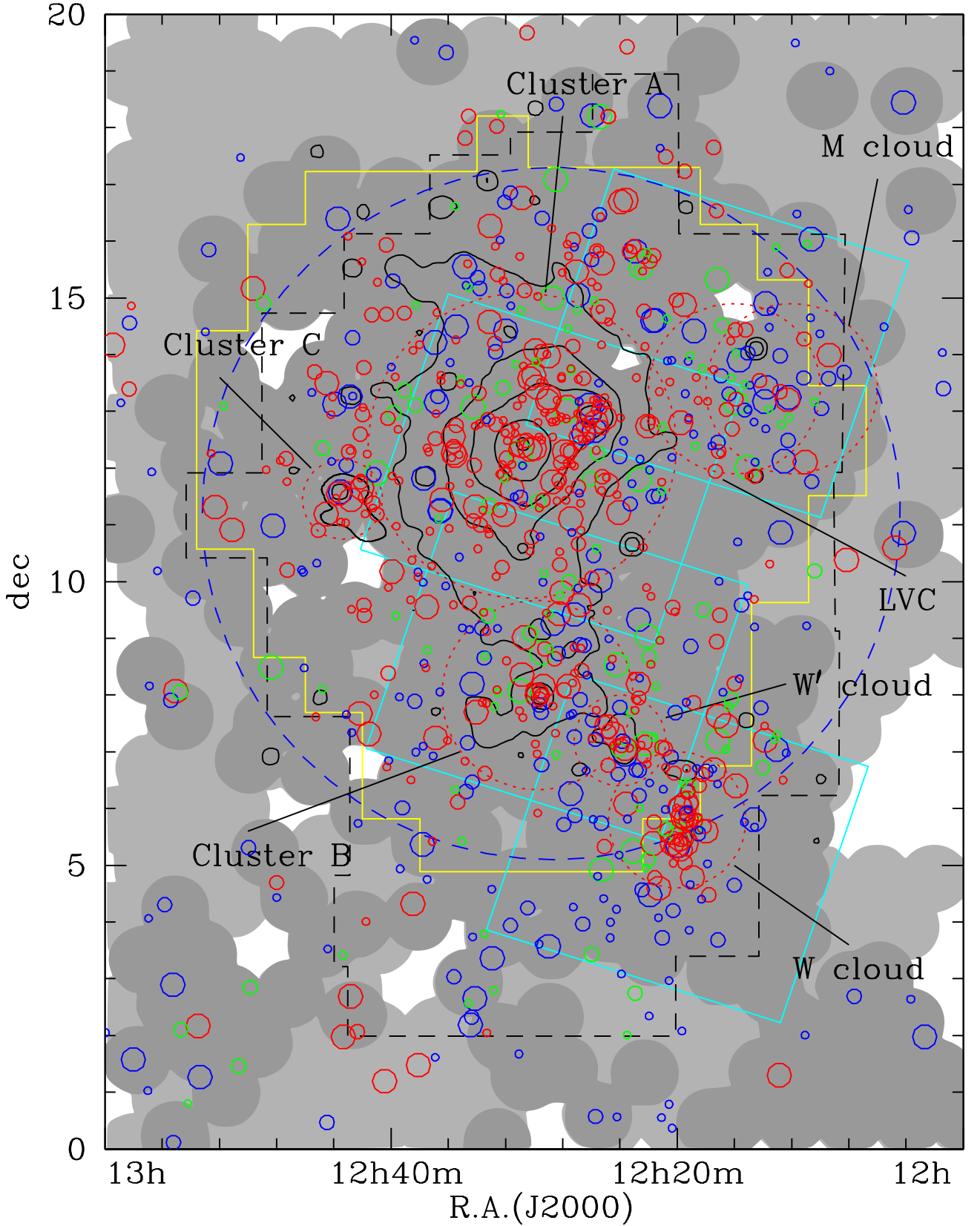}
   \caption{Sky distribution of the GUViCS galaxies with recessional velocity $vel$ $\leq$ 3500 km s$^{-1}$. Red, green, and blue symbols are for galaxies 
   located in the red sequence, green valley, and blue cloud, respectively (see Sect. 5). 
   The size of the symbols is proportional to the stellar mass of the galaxies: big symbols are for galaxies
   with $M_{star}$ $>$ 10$^{9.5}$ M$_{\odot}$, medium size symbols for objects with 10$^{8.5}$ $<$ $M_{star}$ $\leq$ 10$^{9.5}$ M$_{\odot}$ and small size symbols for 
   $M_{star}$ $\leq$ 10$^{8.5}$ M$_{\odot}$.
   The red and blue dashed circles show the identified substructures used in the following analysis. The field is defined as composed by all the galaxies located outside 
   the blue dashed circle and not belonging to any other structure. Light grey indicate shallow (exposure time $<$ 800 sec) NUV \textit{GALEX} fields,
   dark grey deep fields (exposure time $>$ 800 sec).  The footprint of the VCC is shown by the black dashed line, that of NGVS by the yellow solid line, 
   and that of HeViCS by the cyan solid line. The black contours indicate the X-ray diffuse emission of the cluster, from B\"ohringer et al. (1994).}
   \label{angdistmass1}%
   \end{figure*}

To characterise the properties of the ISM of galaxies, our sample of UV selected objects has been also cross-matched with the catalogue of Auld et al. (2013) composed of VCC
galaxies mapped with \textit{Herschel} during the HeViCS survey (Davies et al. 2010). This survey does not cover the full sky region analysed in this work, 
but is limited to the central $\sim$ 84 deg.$^2$ region of the cluster as
depicted in Fig. \ref{angdistmass1}. The survey, which is fairly complete down to $\sim$ 100 mJy at 250 $\mu$m, observed 561 Virgo cluster members
included in our sample, and detected 226 of them\footnote{To be conservative, we consider here as detected galaxies only those with a signal-to-noise larger than 5.}.

Morphological types are taken from the Virgo Cluster Catalogue (VCC, Binggeli et al. 1985) or from its updated revisions (Binggeli et al. 1993). For galaxies outside the Virgo cluster,
morphological types have been determined by us after the visual inspection of the SDSS images. 
As selected, all galaxies have also a redshift measurement necessary to guarantee their membership to the cluster.
Kinematic data are available for a small subsample of galaxies. The ATLAS$^{3D}$ spectroscopic survey (Cappellari et al. 2011a) provides a homogeneous estimate
of the spin parameter measured within the effective radius $\lambda_e$ for 74 bright galaxies (Emsellem et al. 2011), while similar data obtained from long slit spectroscopy 
are also available for 37 dwarf galaxies from the recent compilation of Toloba et al. (2011; 2014). Table \ref{data} summarises 
the completeness of the sample in the different photometric bands.  

In the following analysis we compare the properties of the observed galaxies to those of the Virgo cluster itself locally determined using the X-ray emission of the hot diffuse gas.
The properties of the emitting gas trapped within the potential well of the cluster are here determined using \textit{ROSAT} data from B\"ohringer et al. (1994).

\begin{table}
\caption{Completeness of the sample in the different photometric bands.}
\label{data}
{
\[
\begin{tabular}{ccc}
\hline
\noalign{\smallskip}
\hline
Sample		& N.		& \% \\
\hline
Full sample	& 868		& 100 \\
FUV		& 537		& 62  \\
NUV		& 868		& 100 \\
SDSS(phot)	& 867		& 100 \\
SDSS(spec)	& 575		& 67  \\
\textit{WISE}	& 868(407)	& 100(47)  \\
\textit{Herschel}& 561(226)	& 65(26) \\
HI		& 742(354)	& 85(41) \\
Kinematic	& 111		& 13  \\
\noalign{\smallskip}
\hline
\end{tabular}
\]
Note: in parenthesis the detected galaxies
}
\end{table}

\subsection{The derived parameters}

The multifrequency data in our hands allow us to determine several physical quantities useful for the following analysis.
Distances are determined assuming the mean distance of the different cluster substructures indicated in Table \ref{structures},
as suggested by Gavazzi et al. (1999). For galaxies at the periphery of the cluster, we assume the typical distance of the main body of the cluster (17 Mpc).

UV and optical data are first corrected for dust attenuation. This is done using the prescription of Hao et al. (2011)
based on the relation:

\begin{equation}
{A(NUV) (mag) = 2.5 log[ 1 + 2.26 \frac{L(25 \mu m)}{L(NUV)_{obs}}] }
\end{equation}

\noindent
where $A(NUV)$ is the NUV attenuation (in mag), and $L(25 \mu m)$ and 
$L(NUV)_{obs}$ the observed 25 $\mu$m and NUV luminosities (both expressed in erg s$^{-1}$)\footnote{NUV data, as other photometric data in UV and optical bands, are first corrected for Galactic attenuation 
using the Schlegel et al. (1998) map combined with the Fitzpatrick \& Massa (2007) extinction curve.}, 
which is based on the assumption that the energy absorbed by dust is re-emitted in the far infrared domain. Given that in massive, quiescent late-type galaxies
dust might be heated also by the old stellar population, the attenuation might be slightly overestimated in these systems (Hao et al. 2011). 
The \textit{WISE} data are first multiplied by a factor 1.22, as indicated by Ciesla et al. (2014), to take into account the systematic difference observed 
between the 22 $\mu$m \textit{WISE} and the 24 $\mu$m MIPS bands, assumed to be representative of the 25 $\mu$m band.
The internal attenuation in the other photometric bands, from the FUV-to-the $i$-band, is determined through the relation:

\begin{equation}
{A(\lambda) = A(NUV) \frac{k(\lambda)}{k(NUV)}}
\end{equation}

\noindent
where $k(\lambda)$ is the extinction coefficient determined from the Galactic extinction law of Fitzpatrick \& Massa (2007). 
This determination of the attenuation in the other different bands is a rough estimate since it considers that dust and stars 
of different ages are mixed in a slab model. A more sophisticated correction such as the one proposed by Boselli et al. (2003a)
based on the assumption that the ratio of the thickness of the dusty disc to the stellar disc changes with the mean age of the emitting stars,
thus with the photometric band (sandwich model), is unfortunately impossible because we do not have a measurement of the 
correct inclination of all the galaxies. We notice, however, that the slab approximation is realistic for the UV bands, where
the attenuation is at its maximum and is properly corrected using the prescription given in eq. 2. Any systematic effects in the other bands
should however be minor ($\sim$ 0.1 mag) compared to the colour variations observed among the different galaxy populations ($\sim$ 3 mag). 
The correction for dust attenuation is applied only to late-type systems. \textit{WISE} data are available for most of the 
massive galaxies, while are lacking for a significant fraction of the dwarf star-forming systems (see Appendix A). Given their low metallicity
and dust content, we expect that dust attenuation in these objects is relatively low. We thus do not apply any correction in \textit{WISE}
undetected late-type systems.  \\

Once corrected for attenuation, these data are used to estimate stellar masses. This is done using the prescription of Zibetti et al. (2009)
based on the $i$-band luminosity combined with the $g-i$ colour index. Although this is a standard prescription generally used in the literature,
we recall that it might give erroneous results in galaxies that recently truncated their star formation activity on a very short timescale such
as those analysed in this work. We discuss any possible systematic 
effect in the determination of the stellar mass of galaxies related to environmental effects in Appendix B. \\

The ALFALFA and GOLDMine HI data are used to determine the HI-deficiency parameter, defined as the difference, on logarithmic scale, between the expected and the observed
HI gas mass of each single galaxy (Haynes \& Giovanelli 1984). The expected atomic gas mass is the mean HI mass of a galaxy of a given optical size and morphological type 
determined in a complete sample of isolated galaxies taken as reference. The HI-deficiency parameter has been measured only in late-type galaxies 
using the recent calibrations of Boselli \& Gavazzi (2009).
We consider as unperturbed objects those late-type galaxies with an HI-deficiency parameter $HI-def$ $\leq$ 0.4.\\

Kinematic data are used to differentiate rotationally supported from pressure supported systems. This is done by means of the spin parameter $\lambda$
defined as in Emsellem et al. (2011). Emsellem et al. (2011) identify as fast rotators those objects where $\lambda_e$ $>$ 0.31$\sqrt{\epsilon}$, where $\epsilon$
is the ellipticity of the galaxy\footnote{With respect to the original classification of Emsellem et al. (2011), M60 is here considered as a slow rotator
for the reasons given in section 8.1.}. 



\section{The models}

We compare the observational results to the predictions of models of galaxy evolution specially tailored
to simulate the effects induced by the interaction of galaxies with the hostile cluster environment extensively described in Boselli et al. (2006; 2008a).
The evolution of galaxies is traced using the multizone chemical and spectrophotometric models of Boissier \& Prantzos (2000), updated with an empirically 
determined star formation law (Boissier et al. 2003) relating the star formation rate to the total gas surface densities. These models were shown to
reproduce realistic multiwavelength profiles in comparison to those observed in SING galaxies (Munoz-Mateos et al. 2011).
These models are modified to simulate 
two different effects induced by the interaction of galaxies with the hot intergalactic medium permeating the potential well of the cluster.
In the starvation scenario (Larson et al. 1980; Balogh et al. 2000; Treu et al. 2003), the cluster acts on large scales by removing 
any extended gaseous halo surrounding the galaxy, preventing further infall of such gas onto the disk. The galaxy then becomes anemic 
simply because it exhausts the gas reservoir through ongoing star formation. Our unperturbed model galaxy does not have a hot halo gas.
To reproduce its chemo-spectrophotometric radial gradients, however, the model requires the infall of pristine gas from the surrounding medium. 
This infall is rapid at early epochs in massive objects, while more gradual in time in low-mass systems (Boissier \& Prantzos 2000; Munoz-Mateos et al. 2011).  
Starvation has been simulated just by stopping the infall of gas in the model. We recall that this definition is rather different than the one often adopted in cosmological simulations,
where the hot halo gas is generally instantaneously removed once the galaxy enters the massive dark matter halo of the cluster (e.g. De Lucia 2011). In our model 
starvation is a passive phenomenon, where the gas is only consumed via star formation.\\

The second simulated effect is the ram pressure exerted by the dense intracluster medium ($\rho$ $\sim$ 2 10$^{-3}$ atoms cm$^{-3}$; Boselli \& Gavazzi 2006)
on galaxies crossing the cluster at high velocity ($\sim$ 1000 km s$^{-1}$, Gunn \& Gott 1972). Gas removal induces a quenching of the star formation activity, 
making galaxies redder\footnote{In a ram pressure stripping event, the infall of pristine gas is also stopped.}.
The ram pressure stripping event is simulated by 
assuming a gas-loss rate inversely proportional to the potential of the galaxy, with an efficiency depending on the IGM gas density radial 
profile of the Virgo cluster given by Vollmer et al. (2001). 

Both starvation and ram pressure stripping models have been determined for galaxies with a spin parameter $\lambda$ = 0.05, the typical value for normal
late-type galaxies as those analysed in this work (Mo et al. 1998; Munoz-Mateos et al. 2011), and rotational velocity of 40, 55, 70, 100, 130, and 220 km s$^{-1}$ in order to reproduce 
galaxies spanning a wide range in total mass. For the ram pressure model, we use the stripping efficiency $\epsilon_{0}$ = 1.2 M$_{\odot}$ kpc$^{-2}$ yr$^{-1}$ that reproduces the 
radial profiles of the Virgo cluster galaxy NGC 4569 (Boselli et al. 2006). In Boselli et al. (2006, 2008a) we have shown how the physical properties of galaxies (gas content, 
star formation rate, colours...) change as a function of time and depend on the gas stripping efficiency. Although the effects strongly depend on the assumed value of $\epsilon_{0}$, 
we have adopted the values suggested by these studies. We note that this calibration was made at the present epoch. The relative effect would be smaller at earlier epochs because of a
lower density of the intracluster medium and a lower velocity dispersion within the younger cluster. However, multiple crossing of the cluster occuring every $\sim$ 1.7 Gyr (Boselli \& Gavazzi 2006) 
make the ram pressure stripping process more and more efficient to alter the galaxies properties as time passes (Boselli et al. 2008a).
To mimic starvation, the infall is stopped at different epochs, from the present to 9.7 Gyrs ago (equivalent to $z$ $\sim$ 1.6).
To quantify the effects induced by the cluster environments, all ram pressure stripping and starvation models are compared to those of unperturbed objects of similar
rotational velocity and spin parameter. These models of unperturbed galaxies are indicated as orange filled squares in the following figures.

\section{The colour-stellar mass relation}

\begin{figure*}
   \centering
   \includegraphics[width=18cm]{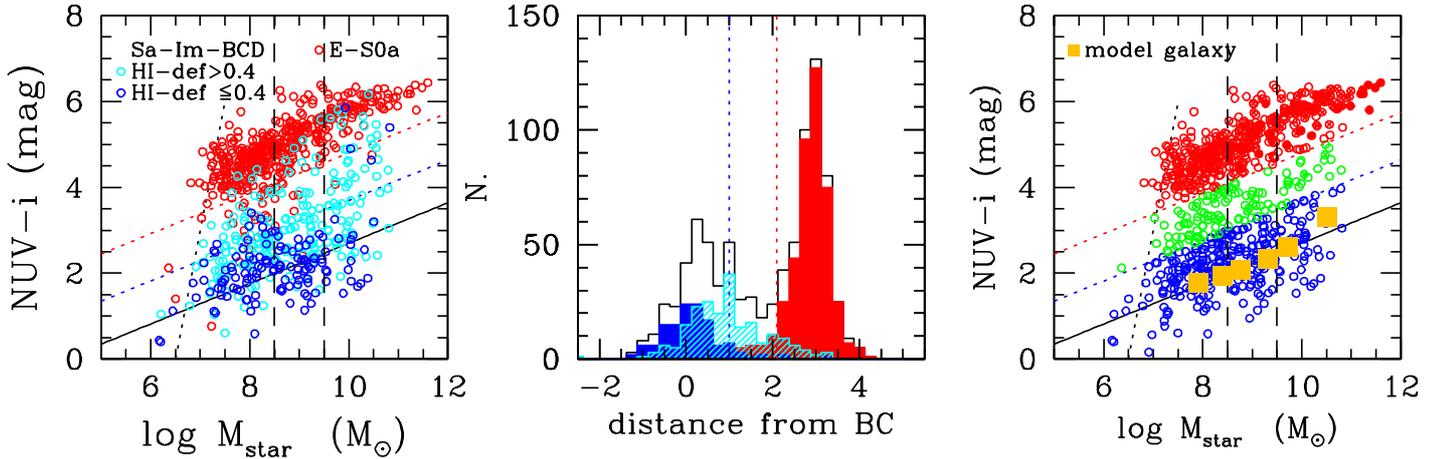}
   \caption{Left: The extinction corrected $NUV-i$ (AB system) vs. $M_{star}$ relation for all galaxies of the sample. 
   Red symbols are for early-type systems (dE-E-S0a), dark and light blue symbols for gas normal ($HI-def$ $\leq$ 0.4) and gas-deficient ($HI-def$ $>$ 0.4)
   late-type galaxies (Sa-Im-BCD). The solid black line shows the typical $NUV-i$ vs. $M_{star}$ relation for model unperturbed late-type galaxies. The black dotted vertical line 
   shows the limit of the GUViCS survey which is more efficient at detecting blue star-forming systems than red objects. The vertical dashed lines show the interval in stellar mass used 
   to define massive, intermediate, and low-mass objects within the sample. The red and blue dotted lines indicate the limit used to define the red sequence and the blue cloud. Centre: 
   Distribution of the colour difference $(NUV-i)$ - $(NUV-i)_{Mod}$ (distance from the blue cloud BC) for the whole sample (black), for early-type galaxies (red), 
   and for HI-normal ($HI-def$ $\leq$ 0.4; dark blue) 
   and gas-deficient ($HI-def$ $>$ 0.4; light blue) late-type galaxies. The vertical red and blue dotted lines indicate the limits used to identify the red sequence and the blue cloud.
   Right: $NUV-i$ vs. $M_{star}$ relation where galaxies are coded according to their belonging to the red sequence (red), green valley (green), and blue cloud (blue). The 
   unpeturbed model galaxies are indicated by orange filled squares. Lines are as in the left panel.
   }
   \label{CMRmass}%
   \end{figure*}

The colour-stellar mass relation has been often used in the literature to identify galaxies at different stages of their evolution in various environments
(Boselli et al. 2008a, 2014; Hughes \& Cortese 2009; Cortese \& Hughes 2009; Gavazzi et al. 2010, 2013a,b). 
This relation is indeed of paramount importance for comparison with models in the study of the formation of the red quiescent galaxy population dominating rich clusters (e.g. Boselli et al. 2008a).
Figure \ref{CMRmass} shows the $NUV-i$ vs. $M_{star}$ relation for the observed galaxies. The $NUV-i$ colour index is sensitive to the relative weight of young stars 
emitting in the UV bands (Kennicutt 1998a; Boselli et al. 2001, 2009) and the bulk of the stellar population dominated by evolved stars in the optical $i$-band.
It is thus a direct tracer of the mean age of the underlying stellar population very sensitive to abrupt variations of the star formation activity as those expected to 
affect cluster galaxies. In the left panel of Fig. \ref{CMRmass} galaxies have different symbols according to their morphological classification and, in late-type objects, to their HI gas content. 
It is evident that, as previously noticed by Hughes \& Cortese (2009), Cortese \& Hughes (2009), and Gavazzi et al. (2013a,b), 
gas-poor late-type galaxies have, on average, redder $NUV-i$ colours than unperturbed galaxies of similar stellar mass.

The $NUV-i$ colour-stellar mass relation has often been used to resolve the red sequence, composed of quiescent early-type systems, from the blue cloud of star-forming, late-type galaxies 
(Gil de Paz et al. 2007) and to identify galaxies in the region between these two sequences, generally called the green valley (e.g. Martin et al. 2007).
Following Cortese \& Hughes (2009), we use this diagram to separate galaxies in the three colour-stellar mass sequences. The red sequence can be easily identified and separated in the diagram 
using the relation $NUV-i$ = 0.47 log$M_{star}$ + 0.1\footnote{This limit corresponds to a typical colour $\sim$ 1.5$\sigma$ bluer than the median $(NUV-i)$ - $(NUV-i)_{Mod}$
colour of early-types, where $\sigma$ is the typical dispersion of the red sequence.}. The separation between the blue cloud and the green valley is less direct in particular in this sample which includes 
a large fraction of perturbed objects.
To select galaxies in the three sequences, we first define the typical $NUV-i$ vs. $M_{star}$ relation expected for field systems using our models of galaxy evolution for unperturbed objects (right panel;
$(NUV-i)_{Mod}$ = 0.47 log$M_{star}$ - 2.0). Curiously the slope of the relation corresponds to that determined to separate the red sequence. To define the dynamic range in the $NUV-i$ colour
typical of unperturbed blue cloud galaxies, we plot in Fig. \ref{CMRmass} the distribution of the colour difference $(NUV-i)$ - $(NUV-i)_{Mod}$ or, in other words, 
the distance in colour from the blue cloud at a given stellar mass, for HI-normal and HI-deficient ($HI-def$ $\leq$ and $>$ 0.4) galaxies (central panel). 
Galaxies with an HI-deficiency parameter $HI-def$ $\leq$ 0.4 are here taken as representative of the typical unperturbed field population. 
Figure \ref{CMRmass} clearly shows that the distribution in colour of the HI-normal late-type systems is symmetric and peaked around the $NUV-i$ vs. $M_{star}$ relation 
drown by the unperturbed model galaxies. It drops to $\simeq$ 0 at $(NUV-i)$ - $(NUV-i)_{Mod}$ = $\pm$ 1 mag, while that of HI-deficient cluster galaxies is, on average redder. 
We thus define the limit between the blue cloud and the green valley with the relation $NUV-i$ = 0.47 log$M_{star}$ - 1.0. The slope of the $NUV-i$ vs. $M_{star}$ relation 
described by the model galaxies used here to select objects belonging to the blue cloud is steeper than the one determined using different sets of data by Wyder et al. (2007) or Cortese et al. (2009). 
These definitions, as every definition of red sequence, green valley, and blue cloud galaxies,
are still quite arbitrary. We recall, however, that we use these definitions to identify galaxies according to the mean age of their underlying stellar
population and study their relative distribution within the cluster (centre vs. periphery, in the different substructures or as a function of galaxy density).
Relative measurements, at least at the first order, are not sensitive to the adopted definitions of the three different sequences. Indeed, the main results of this work do not change 
whether different definitions are used.

\section{The galaxy distribution within the cluster}

\subsection{Identification of the substructures on the plane of the sky}

Once identified according to their spectrophotometric properties, we can use the complete coverage of the GUViCS survey to see how 
the different kinds of galaxies are distributed within the various known substructures composing the Virgo cluster and its surrounding regions. To do that, 
we plot in Fig. \ref{angdistmass1} the distribution on the plane of the sky of all the galaxies of the sample with different symbols to identify galaxies
in the three colour sequences in three different bins of stellar mass. We also plot the contours of the X-ray emitting gas showing the distribution
of the hot and dense intracluster medium.


\noindent
Thanks to the large number of galaxies with spectrophotometric data and spectroscopic information within the surveyed region, 
Figure \ref{angdistmass1} can be used to identify the different substructures of the Virgo cluster, originally defined by Binggeli et al. (1987, 1993). 
An overdensity of galaxies is indeed observed 
on the main peak of the X-ray emission of the cluster, centered on M87. This structure, which is the main structure of the whole cluster, 
is generally called Virgo cluster A. South of this region, at R.A. $\sim$ 12h30m and dec $\sim$ 8$^o$,
there is another condensation of galaxies associated with a second peak of the X-ray emission. This substructure is generally called Virgo cluster B,
centered on the giant elliptical M49. A third peak of density is present east of Virgo cluster A, at R.A. $\sim$ 12h45m and dec $\sim$ 11.5$^o$,
again associated with a peak of X-ray emission, centered on the elliptical galaxy M60 (Virgo cluster C). There are three other obvious galaxy overdensity regions, 
one associated with a hot diffuse gas overdensity, at R.A. 12h24m and dec 7.2$^o$ (W' cloud), the other two at R.A. 12h20m and dec 5.8$^o$ (W cloud) and at
R.A. 12h12m and dec 13.4$^o$ (M cloud). East of the M cloud there is a second structure easily identifiable in the velocity space, generally called the Low Velocity Cloud (LVC; Hoffman et al. 1989).

\subsection{Identification of the substructures in the velocity space}

\begin{figure}
   \centering
   \includegraphics[width=9cm]{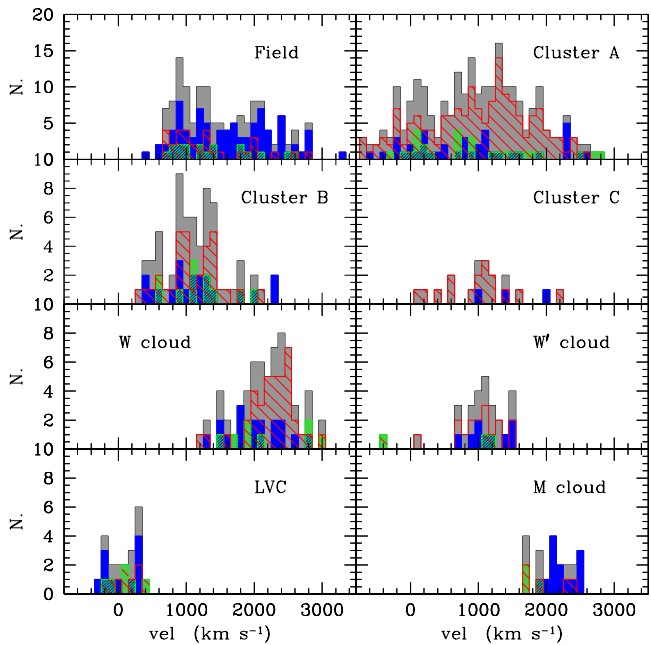}
   \caption{Velocity distribution of galaxies within the different cluster substructures and in the field. The grey histogram shows the distribution of all galaxies, 
   the red, green, and blue histograms those of objects belonging to the red sequence, the green valley, and the blue cloud, respectively. }
   \label{velocita}%
   \end{figure}

These structures, characterised by different mean recessional velocities, are known to be at different distances (Binggeli et al. 1987, 1993; Gavazzi et al. 1999).
We thus use velocity constraints as indicated in Table \ref{structures} to assign the membership of galaxies to the different substructures, as 
originally done by de Vaucouleurs (1961), Ftaclas et al. (1984), Binggeli et al. (1987), and Hoffman et al. (1989). 
For the purpose of the present work, we identify as members of Virgo cluster A and B those galaxies with an angular distance smaller than half the virial radius of these two 
structures measured by McLaughlin (1999) and Ferrarese et al. (2012). The choice of half a virial radius is taken to avoid the quite uncertain identification of galaxies in possible overlapping regions in
the two substructures. For the remaining substructures we take arbitrary values for limiting the angular distance from the overdensity peak. This choice is dictated by the fact that
here the identification of virial entities using analytic prescriptions such as the one proposed by Finn et al. (2005) might be not appropriate
given the unrelaxed nature of these substructures. The resulting distribution of galaxies belonging to the different substructures in the velocity space is shown in Fig. \ref{velocita}.

\noindent
Galaxies located outside a circular region centered at R.A. 12h29m and dec 11.2$^o$ of radius 6.1 degrees (corresponding to $\simeq$ 1.8 Mpc at the typical distance of Virgo here assumed to be 17 Mpc) 
and not belonging to the other substructures are here considered as field objects and will be used as reference sample in the following analysis. 
We recall, however, that these objects are not genuine field galaxies just because they are located at the periphery of the cluster. This limiting region, indeed, roughly corresponds
to the virial radius of cluster A ($R_{vir}(A)$ $\simeq$ 1.6 Mpc) and might thus be still affected by the cluster environment. Indeed, their velocity distribution is peaked at $\sim$ 1000 km s$^{-1}$,
i.e. at the mean recessional velocity of cluster A. 

\begin{table*}
\caption{The properties of the different cluster substructures (see Fig. \ref{angdistmass1}).}
\label{structures}
{
\[
\begin{tabular}{ccccccccccc}
\hline
\noalign{\smallskip}
\hline
Substructure	& R.A.(J2000)	& dec	& Radius	& velocity range	& Dist$^a$	& $<vel>$	& $\sigma$	& N. 	& $\rho$ $^b$	&Central galaxy \\
		& $^o$		& $^o$	& $^o$		& km s$^{-1}$		&Mpc		& km s$^{-1}$	&  km s$^{-1}$	&	& N.gal. Mpc$^{-3}$ &	\\
\hline
Cluster A	& 187.71	& 12.39	& $<$2.692$^c$	& $<$ 3500		& 17		& 955		& 799		& 234	& 110	& M87	\\
Cluster B	& 187.44	& 8.00	& $<$1.667$^c$	& $<$ 3500		& 23		& 1134		& 464		& 66	& 53	& M49	\\
Cluster C	& 190.85	& 11.45	& $<$0.7	& $<$ 3500		& 17		& 1073		& 545		& 19	& 507	& M60 	\\
W cloud		& 185.00	& 5.80	& $<$1.2	& 1000$<$ $vel$ $<$3500	& 32		& 2176		& 416		& 66	& 52	& NGC4261	\\
W' cloud	& 186.00	& 7.20	& $<$0.8	& $<$ 2000		& 23		& 1019		& 416		& 28	& 202	& NGC4365	\\
M cloud		& 183.00	& 13.40	& $<$1.5	& 1500$<$ $vel$ $<$3500	& 32		& 2109		& 280		& 21	& 9	& NGC4168	\\
LVC cloud	& 184.00	& 13.40	& $<$1.5	& $vel$ $<$400		& 17		& 85		& 208		& 21	& 57	& NGC4216	\\
Field		& 187.20	& 11.20	& $>$6.1	& $<$ 3500		& 17		& 1537		& 635		& 138	& 0.05$^d$& -		\\
\noalign{\smallskip}
\hline
\end{tabular}
\]
Note: Galaxies in the overlapping regions satisfying the membership criteria of two different structures are assumed to be members of the smallest structure.\\ 
$a$: mean distance of each single substructure taken from Gavazzi et al. (1999).\\
$b$: mean density assuming a spherical geometry of the substructure.\\
$c$: radius corresponding to half of the virial radius, from McLaughlin (1999) and Ferrarese et al. (2012). \\
$d$: the mean density at the periphery of the cluster is probably underestimated because there the \textit{GALEX} observation are, on average,
less deep than in the inner $\sim$ 100 deg$^2$.}
\end{table*}

\subsection{Determination of the galaxy density}

The data in our hand can be used to estimate the mean galaxy density around each single object. This exercise, however, is made difficult
by the elongated 3D-structure of the cluster. Indeed, Virgo is composed of different substructures overlapping on the line of sight and 
located at different distances. Unfortunately an accurate determination of the distance is available only for a small fraction of the targets.
Given the large velocity dispersion within the cluster, the recessional velocity cannot be taken as a distance tracer through the Hubble relation.
We thus decided to estimate mean surface densities rather than volume densities of galaxies. To do that, we cut the Virgo cluster region in the velocity space 
to separate galaxies belonging to the different substructures as indicated in Table \ref{structures}. We then apply two different methods to calculate 
the local density of galaxies around each object of the sample.\\

The first method consists in counting the number of galaxies within a cylinder of radius 0.2 Mpc (calculated assuming the mean distance 
of each single substructure) centered on each single galaxy and of depth corresponding to the 
velocity range indicated in Table \ref{structures}. For simplicity, all "field" galaxies and galaxies not included in any of the regions 
indicated in Table \ref{structures} with recessional velocity $<$ 3500 km s$^{-1}$ are assumed at 17 Mpc. This strong assumption probably induces an overestimate 
of the galaxy density in the periphery of the cluster, where a significant fraction of galaxies might be in Hubble flow. We recall, however, 
that all galaxies observed within the GUViCS survey with a recessional velocity $<$ 3500 km s$^{-1}$ are located within $\simeq$ 2 virial radii of cluster A 
and can thus be considered as Virgo members, as indicated by the velocity distribution shown in Fig. \ref{velocita}. The choice of such a small 
radius for the cylinder is dictated by the fact that we want to estimate density variations on relatively small scales, i.e. on scales comparable to those of the 
smallest substructures already identified in Virgo.\\

The second method is the Voronoi tessellation method. For each galaxy we first define the Voronoi cells as the polygonal cells centered on the galaxy 
and enclosing all the surrounding empty space closest to that point. This method has proven to be very efficient in determining galaxy densities in different environments
(Platen et al. 2011; Scoville et al. 2013, Cybulski et al. 2014) and even in finding bound structures like galaxy groups (Marinoni et al. 2002; Gerke et al. 2005; Cucciati et al. 2010). 
The local density around the galaxy is then simply given by the inverse of the area of the Voronoi cell.
To avoid projection effects, we adopt the same cut in the velocity space as those adopted for the cylinder method. 

Both methods are applied only to those galaxies observed in deep
observations (those located within the dark grey regions in Fig. \ref{angdistmass1}) to
avoid systematic effects in the density estimate. Indeed, in the outer
regions, where the \textit{GALEX} observations are shallower, the mean density
might results lower just because the faintest galaxies are not detected.
In the case of the cylinders, this is taken into account dividing the
local density around each galaxy by the fraction of the volume of the
cylinder that falls within the area covered by the deep observation. This
method has already been effectively used to correct for boundary effects
(see eg. Cucciati et al. 2006). In the case of the Voronoi tessellation, we
did not try to correct for boundary effects given by the the use of the
deep data only. This results in a possible underestimate of the local
density for galaxies residing in the Virgo outskirts. We decided not to
correct the Voronoi local density because the Voronoi tessellation is an
adaptive method that strongly depends on galaxy relative position, and it
is harder to correct for boundary effects not knowing where the closest
galaxy lies. Nevertheless, as explained below, we are more interested in
contrasting low and high densities than determining the exact density
value. Moreover, we find very similar results when using the density
computed in cylinders and the one computed with the Voronoi tessellation.

The determination of the local density around galaxies using either method strongly depends on several properties of the selected sample, such as
its sensitivity in surface brightness and its completeness in total magnitude and redshift. Since these properties strongly depend 
on the adopted selection criteria used to define the sample, we decided to measure for each galaxy a density contrast.
This is done as in Gavazzi et al. (2010). We first define the local density of the field: 

\begin{equation}
{\rho_{field} = \frac{N_{field}}{Area_{field}}}
\end{equation}

\noindent
where $N_{field}$ is the number of field galaxies in $Area_{field}$, the field region defined in Table \ref{structures}, with \textit{GALEX} exposures 
in the NUV band $>$ 800 seconds. This limit in the integration time secures the completeness of the sample down to stellar masses of $\simeq$ 10$^8$ M$_{\odot}$.
The density contrast measured at the position of each single galaxy is then defined as:

\begin{equation}
{\Delta \rho = \frac{\rho}{\rho_{field}} -1}
\end{equation}

\noindent
where $\rho$ is the local density around each galaxy estimated either with the cylinder filter or the Voronoi tassellation.
Because of this definition, the density contrast of galaxies in the outskirts of the cluster is $\Delta \rho$ $\simeq$ 0. As shown in Fig. \ref{angdistmass1}, 
$\rho_{field}$ is determined mainly using galaxies located in the southern extension of the Virgo cluster. This is known to be an overdense region with respect to the typical field.
The density contrast $\Delta \rho$ determined with equation 4 gives thus just a contrast of density within the observed region and certainly underestimates 
the real density contrast between the different structures of the cluster and the general field. This, however, is sufficient for the purpose of this work 
since we are interested in quantifying relative differences in local density over the surveyed region rather than determining absolute values.

   \begin{figure*}
   \centering
   \begin{tabular}{c}
   \includegraphics[width=0.4\textwidth]{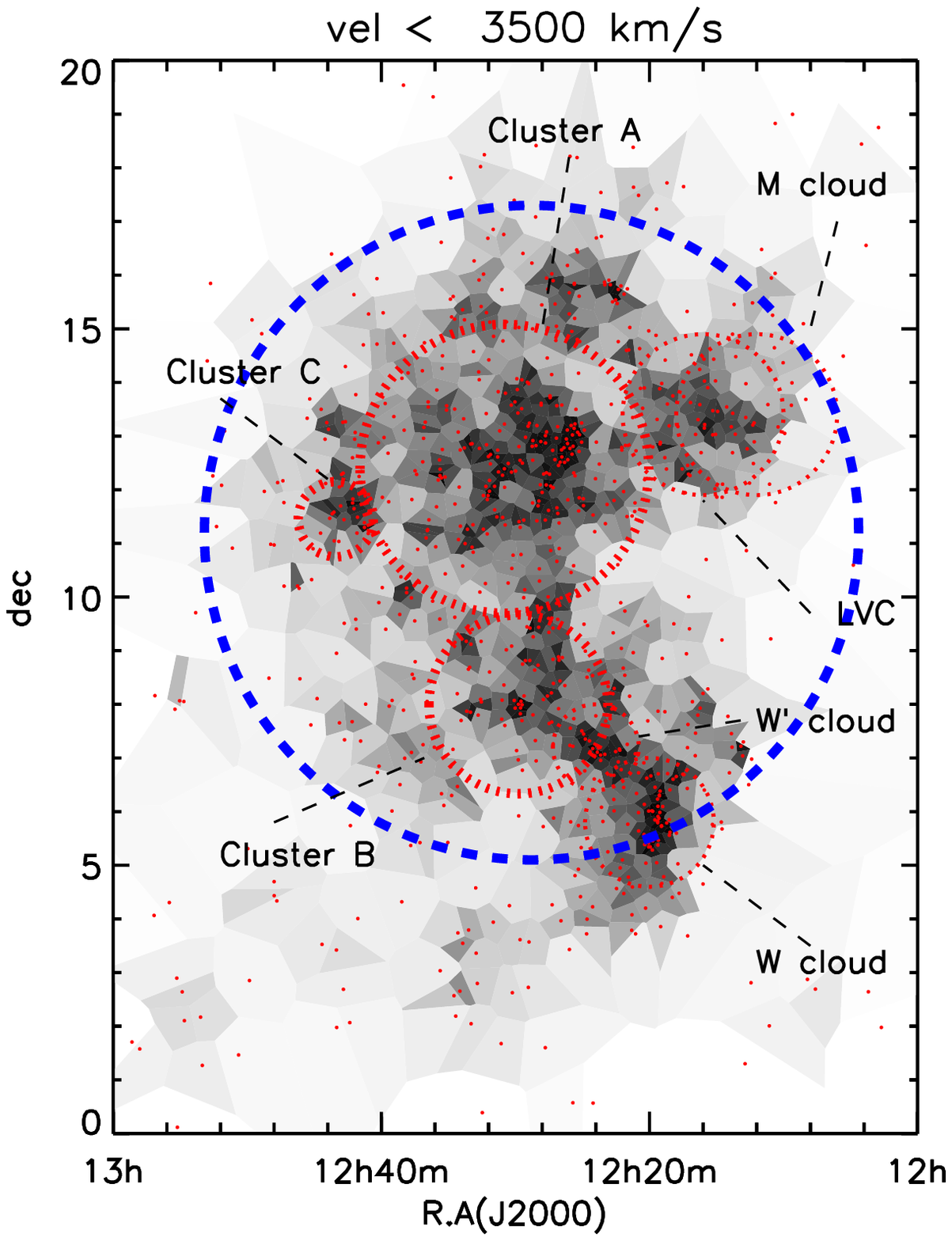}
   \includegraphics[width=0.4\textwidth]{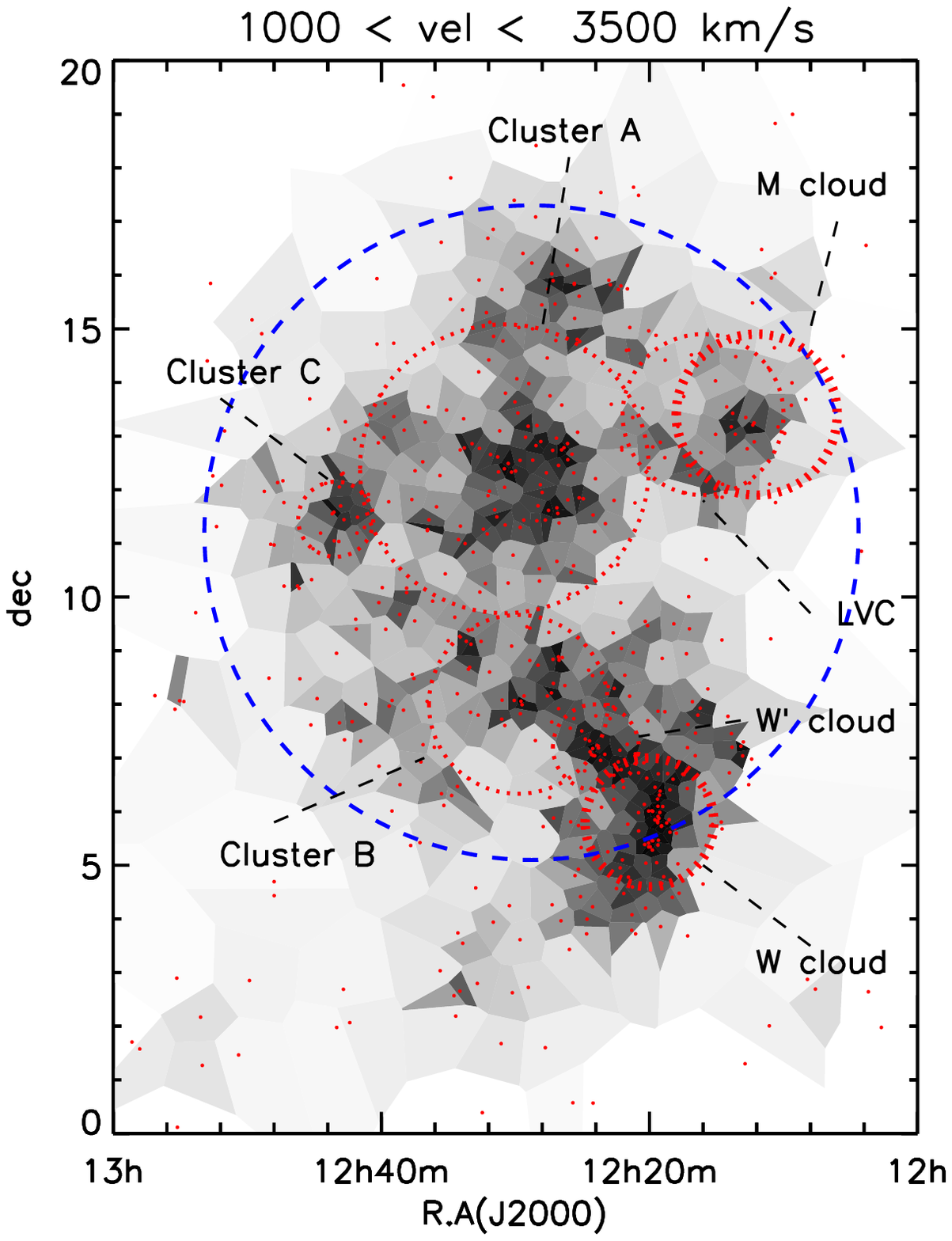}\\
   \includegraphics[width=0.4\textwidth]{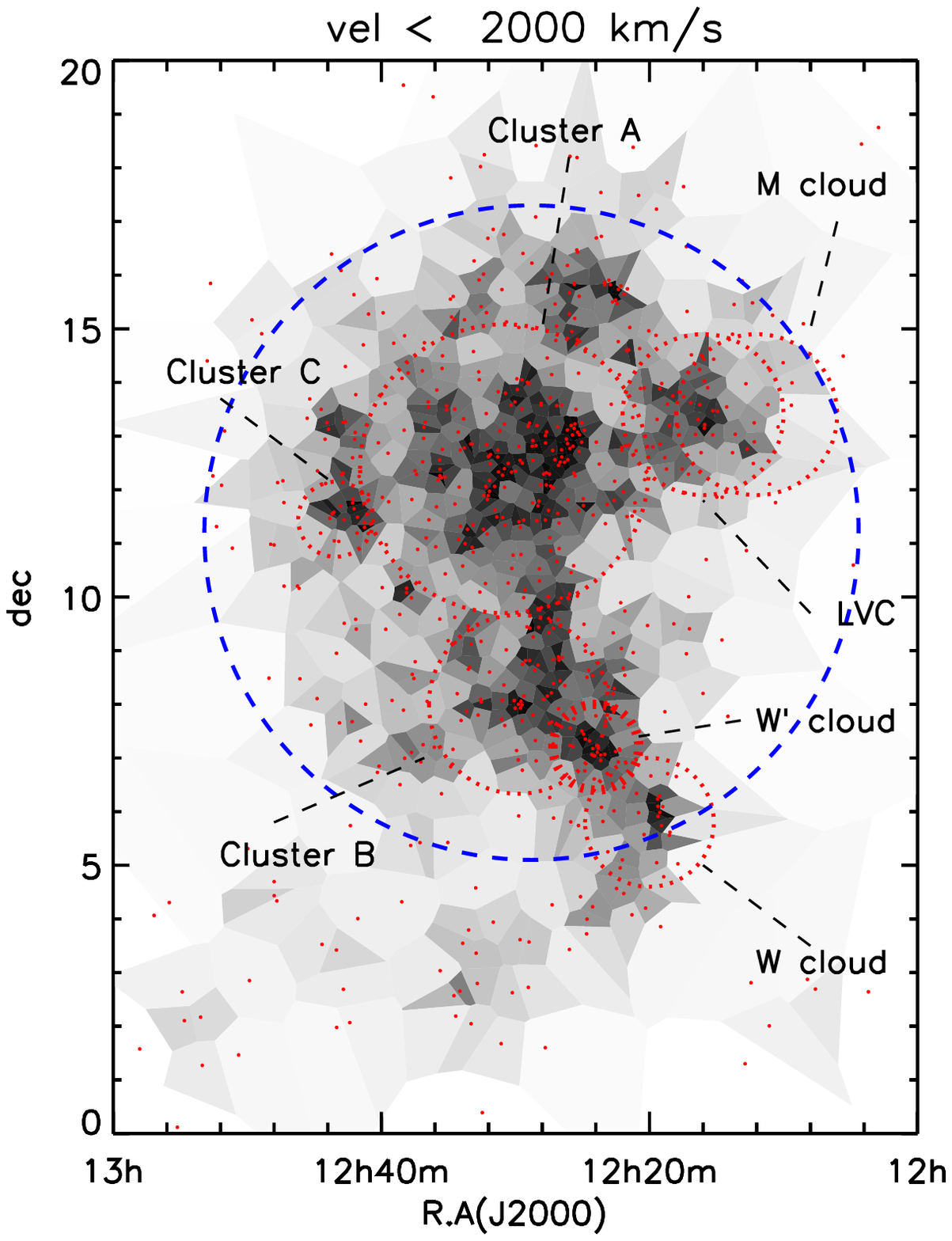}
   \includegraphics[width=0.4\textwidth]{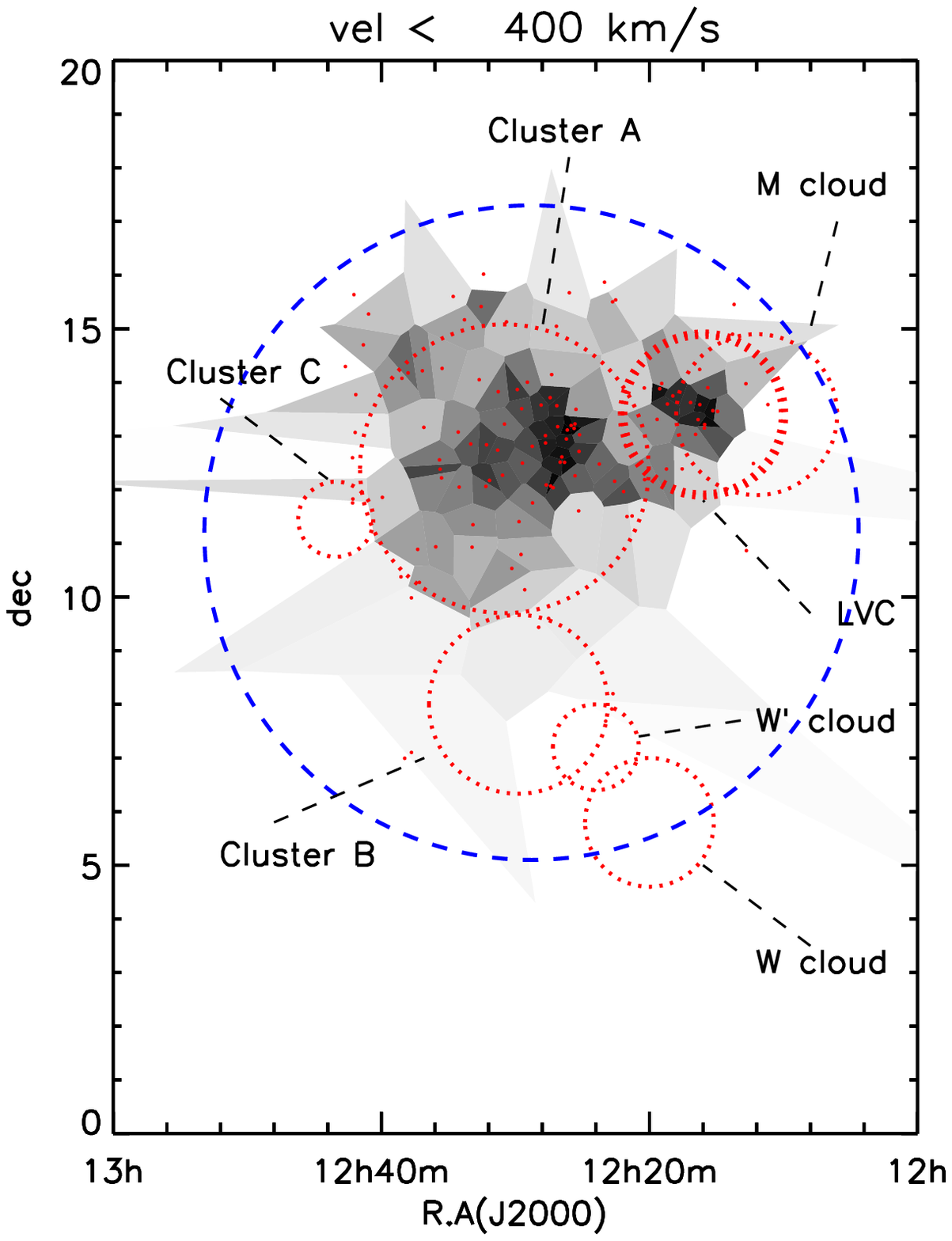}\\
   \end{tabular}
   \caption{The Voronoi tessellation of the Virgo cluster region, with darkness increasing according to the local density. The four plots show the Voronoi tessellation
   done using the whole sample of galaxies with recessional velocity $<$ 3500 km s$^{-1}$ (upper left), and in three subsamples in the velocity space: 1000 $<$ $vel$ $<$ 3500 km s$^{-1}$ (upper right),
   $vel$ $<$ 2000 km s$^{-1}$ (lower left), and $vel$ $<$ 400 km s$^{-1}$ (lower right). Red dots indicate the galaxies.
   The different cluster substructures and the region defining the field are identified with red-dotted and blue-dashed 
   circles. Thicker contours are used whenever the structure has been identified within that velocity range, with the only exception of the M cloud (see Table \ref{structures}). }
   \label{voronoi}%
   \end{figure*}

The comparison of the estimate of the density contrast determined using the cylinder and the Voronoi tessellation method gives very consistent results. The Voronoi technique 
produces a dynamical range of densities much larger than the use of cylinder with a  fixed radius, because it is an adaptive method. In this way, the highest densities are
computed on smaller scales than with the cylinders (even with a fixed radius of 0.2 Mpc), resulting in a much higher density contrast.
Since we want to explore variations of different physical parameters
as a function of the environment, we decided to use the Voronoi density contrast in the future analysis to have the largest possible range in the parameter space.
We check, however, that the main conclusions of this work are robust vs the use of the two different density estimators.
The Voronoi method has also the advantage that it does not depend on any assumption on the size of the adopted cylinder. We recall, however, that it might underestimate 
the local density of galaxies at the boundary regions of the Virgo cluster.
Figure \ref{voronoi} shows the Voronoi tessellation of the Virgo cluster region done using the whole sample of galaxies with redshift $<$ 3500 km s$^{-1}$ and those done in different intervals 
of recessional velocity. All the different substructures previously identified are clearly visible in these figures. That is also the case for the low velocity cloud (LVC), clearly evident 
in the Voronoi tessellation plot done for galaxies with recessional velocity $vel$ $<$ 400 km s$^{-1}$.

\section{The analysis}

\subsection{The 2-D distribution of galaxies within the cluster}

Besides helping us in the identification of the different cluster substructures, Fig. \ref{angdistmass1} clearly shows that galaxies
belonging to the red sequence are preferentially located in the high density regions. The only exception are the M and LVC clouds, where red sequence, 
green valley, and blue cloud objects are well mixed. This evidence is a further confirmation of the well known morphology segregation effect (e.g. Dressler 1980; Whitmore et al. 1993)
that, in the Virgo cluster, is known to extend to the dwarf population (Binggeli et al. 1988). Figure \ref{angdistmass1} also shows that
the highest concentrations of quiescent systems roughly correspond to the peaks of the X-ray emission, where the density of the  
intracluster medium is at its maximum (Schindler et al. 1999)\footnote{Schindler et al. (1999), however, noticed that while the X-ray emission of the cluster is peaked on M87, 
that of galaxies is $\sim$ 1 deg. north-west of M87 in the direction of M86.}. The X-ray map of B\"ohringer et al. (1994) does not extend below 6$^o$ in declination, 
we thus do not know whether the W cloud is also associated to a peak in the X-ray emission. In other words, the X-ray emission is a good tracer of the potential well of the 
different cluster substructures.

\subsection{The colour-stellar mass relation within the substructures}

   \begin{figure*}
   \centering
   \includegraphics[width=18cm]{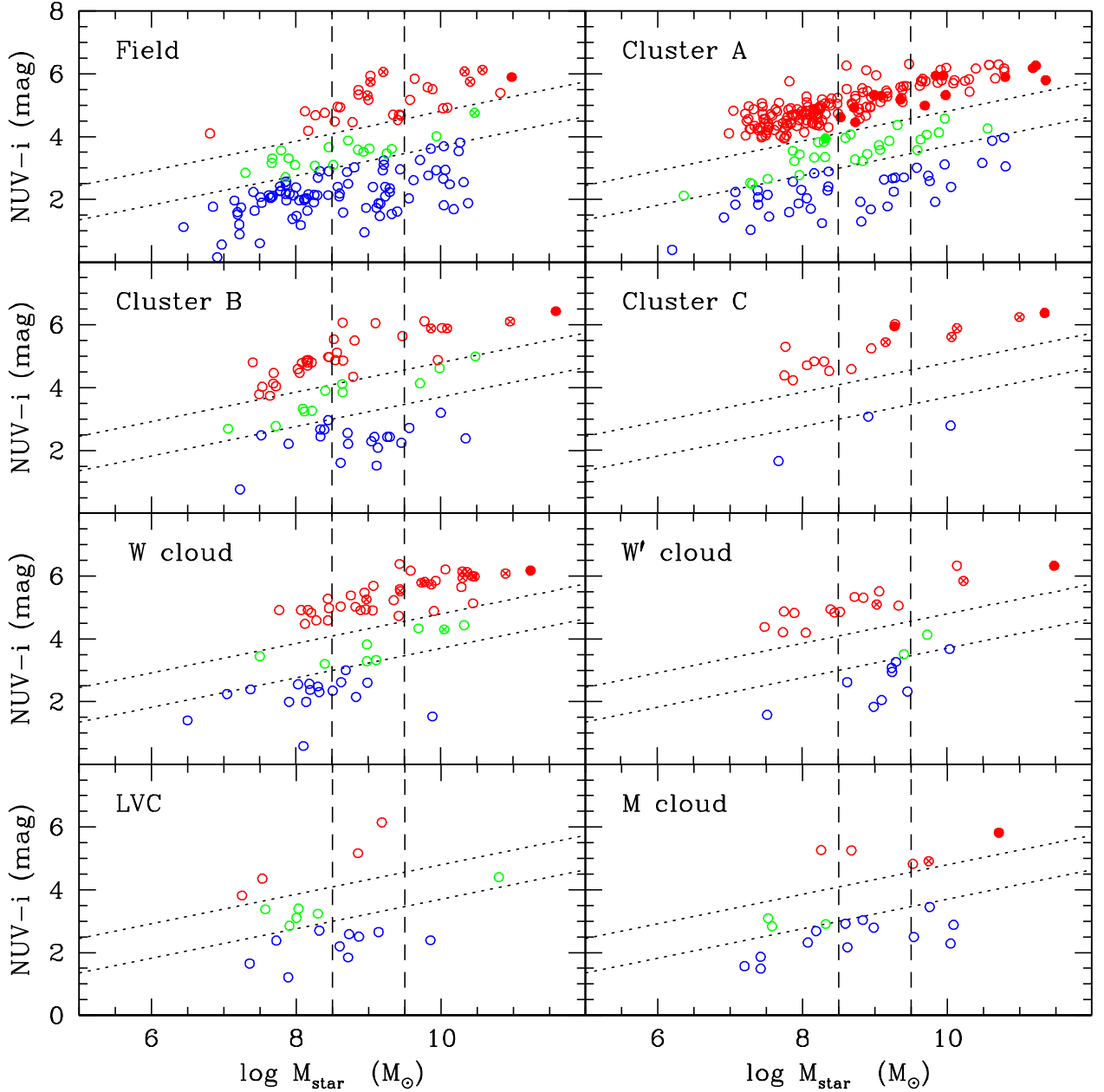}
   \caption{The extinction corrected $NUV-i$ (AB system) vs. $M_{star}$ relations for galaxies
   selected according to their substructure membership as defined in the text. Red, green, and blue colours are for galaxies 
   in the red sequence, green valley, and blue cloud, respectively.
   Filled symbols are for slow rotators, crosses for fast rotators. The vertical dashed lines show the interval in stellar mass used 
   to define massive, intermediate, and low-mass objects within the sample.
   }
   \label{CMRmassall}%
   \end{figure*}

Another way to study the dependence of the mean spectrophotometric properties of galaxies as a function of environment is to 
determine their $NUV-i$ vs. $M_{star}$ relations separately for the different cluster substructures (Fig. \ref{CMRmassall}). 
Figure \ref{CMRmassall} confirms the claim of the previous section, i.e. that all the identified regions, with the exception of the M and LVC clouds,
are dominated by quiescent red galaxies, while galaxies at the periphery of the cluster (field) are mainly late-type systems. It also shows that 
the most massive galaxies ($M_{star}$ $\gtrsim$ 10$^{11}$ M$_{\odot}$), all red early-types with the exception of the spiral NGC 4216 in the LVC, are the dominant galaxies in all 
the substructures, independently from their size or number of objects. Again, the M cloud could be considered as an exception because 
its most massive object, NGC 4168, has a stellar mass of only $M_{star}$ = 10$^{10.73}$ M$_{\odot}$. Among the analysed substructures, we recall that
the M cloud is the one with the lowest galaxy density (Table \ref{structures}). On the contrary, galaxies with 
stellar masses $M_{star}$ $\gtrsim$ 10$^{11}$ M$_{\odot}$ are lacking in the field.

At the faint end of the colour-stellar mass relation, for $M_{star}$ $\lesssim$ 10$^{8}$ M$_{\odot}$, Figure \ref{CMRmassall} also shows a systematic difference in the galaxy 
population between the field and the densest regions, in particular Virgo cluster A. Indeed, in this stellar mass range the core of the cluster 
is dominated by red quiescent dwarf ellipticals while in the field these objects are totally lacking. The other substructures, with the exception
of the M and LVC clouds, seem to share the properties of cluster A (the observed small shift in the limiting stellar mass of the different subsamples
is due to their higher distance, see Table \ref{structures}). The observed difference with the field, on the other hand, can be partly due to a selection bias
that favors the detection of blue objects in the periphery of the cluster where the survey is shallower than in the core (see Fig. \ref{angdistmass1}).\\

   \begin{figure*}
   \centering
   \includegraphics[width=0.4\textwidth]{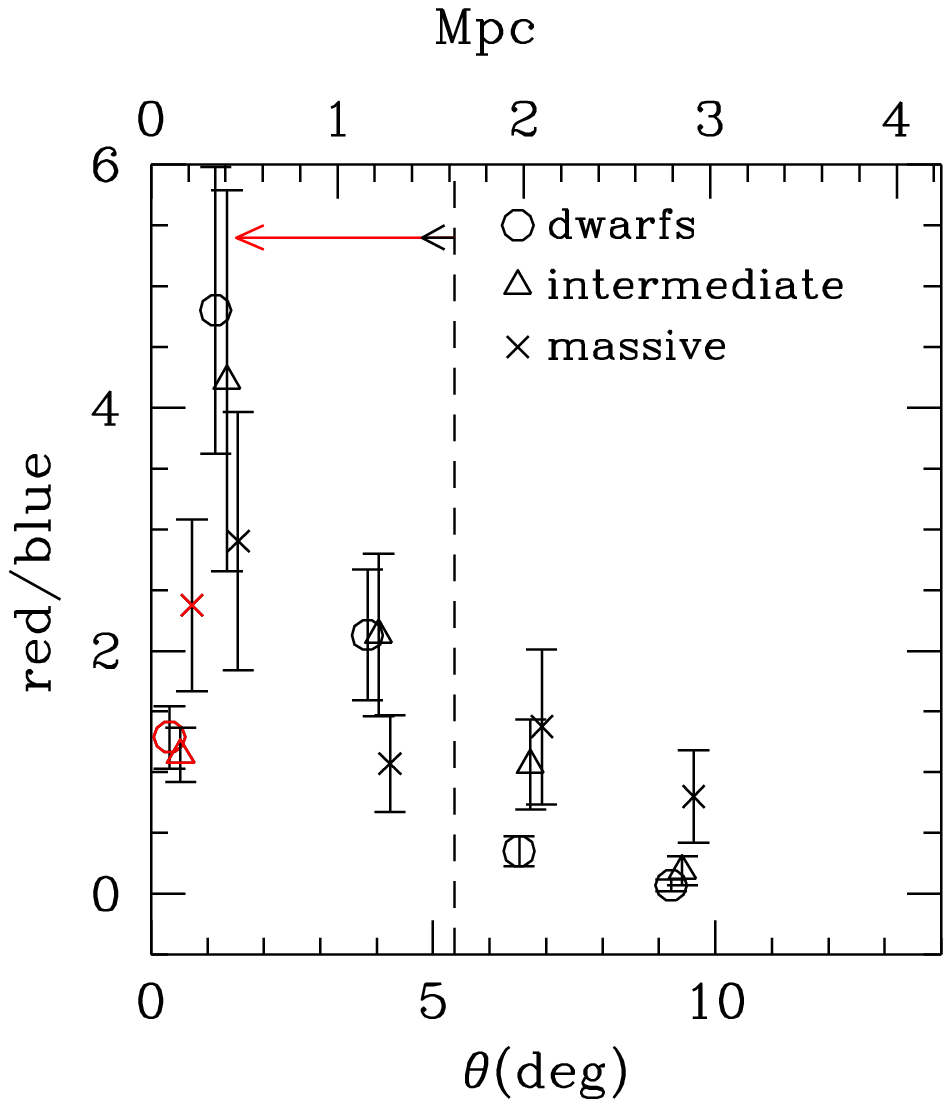}
   \includegraphics[width=0.42\textwidth]{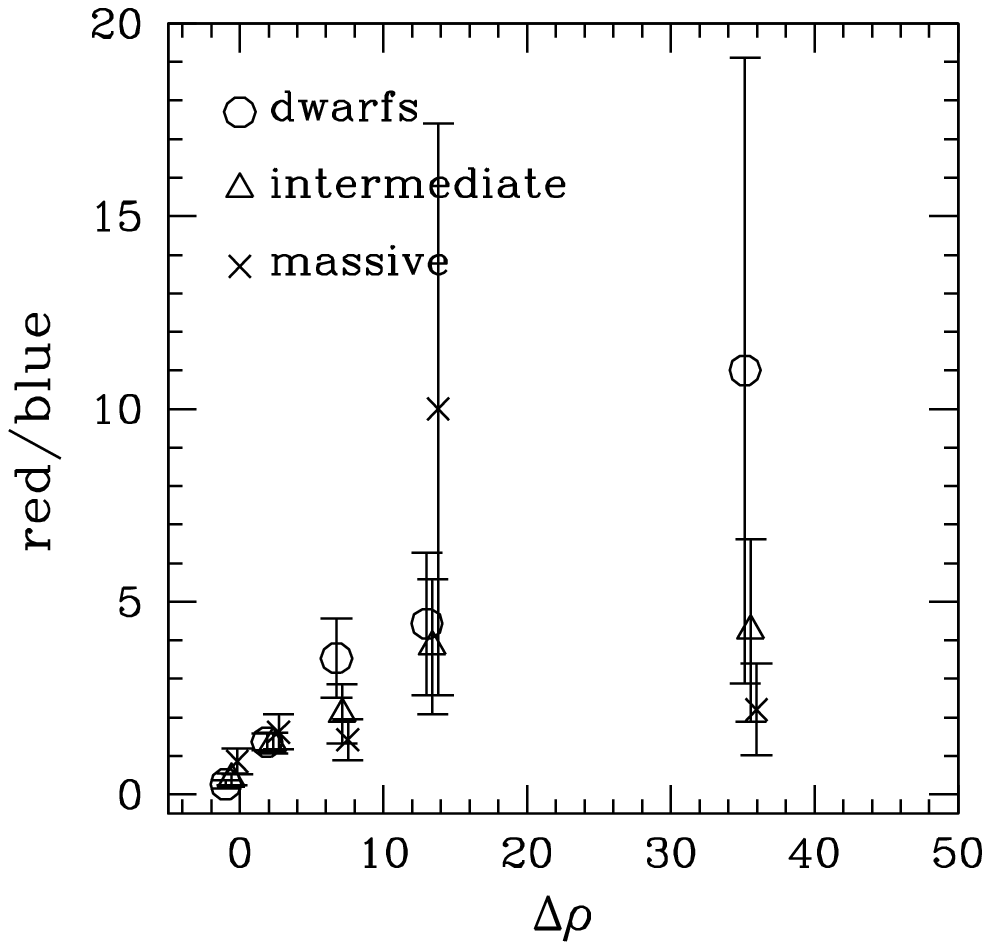}
   \caption{Left: the relationship between the red-to-blue galaxy fraction and the angular distance from the cluster centre for massive (crosses), intermediate (triangles)
   and low-mass (circles) galaxies belonging to cluster A or to the field (black symbols), measured in bins of half virial radii (crosses and circles are slightly shifted along the X-axis 
   to avoid confusion in the Poisson error estimate). Red symbols give the mean values 
   for the other Virgo cluster substructures (clusters B, C and M, W, W', and LVC clouds) at their mean angular distance determined in Mpc.
   The black dashed vertical line indicates the virial radius of cluster A.
   The black and red horizontal arrows show the mean distance covered by dwarf galaxies within the cluster during the time that they need to become HI-deficient (black) 
   and red (red) after the quenching of their star formation activity, as predicted by our models. Right: the relationship between the red-to-blue galaxy fraction and the density contrast.}
   \label{frazione}%
   \end{figure*}

We can also study the radial variation within the cluster of the mean spectrophotometric properties of galaxies 
by plotting the red-to-blue galaxy ratio as a function of the angular distance from the core of Virgo cluster A ($\theta$) and the mean value for the other
cluster substructures (cluster B and C, plus W, W', M and LVC clouds), and of the density contrast ($\Delta \rho$)
for galaxies in three different bins of stellar mass (Fig. \ref{frazione}).
Figure \ref{frazione} clearly shows that the radial variation of the red-to-blue galaxy ratio from the core of cluster A to $\sim$ 2 virial radii is small
in massive systems (a factor of $\sim$ 3), while important in intermediate mass and in dwarf galaxies ($\gtrsim$ 20). This trend is even 
stronger than the one observed by Haines et al. (2006a,b) in other nearby clusters.
Only part of this trend can be due to selection biases (the outer regions are undersampled by \textit{GALEX} at a sensitivity of the MIS).
Indeed, these radial variations are already strong if limited to the intermediate stellar mass range, where the survey is expected to be fairly complete 
even at the periphery of the cluster. It is also clear if limited to the inner $\sim$ 1 virial radius, with no restrictions on the stellar mass, where the survey is homogeneously complete 
at the depth of the MIS. A similar trend, although more scattered, is observed when the red-to-blue galaxy ratio is plotted vs. the density contrast. Curiosily, the mean
fraction of red-to-blue galaxies in the other cluster substructures is significantly lower than the one observed in the core of cluster A and does not significantly change with galaxy mass.

\subsection{The colour-stellar mass relation as a function of the density contrast}

   \begin{figure*}
   \centering
   \includegraphics[width=18cm]{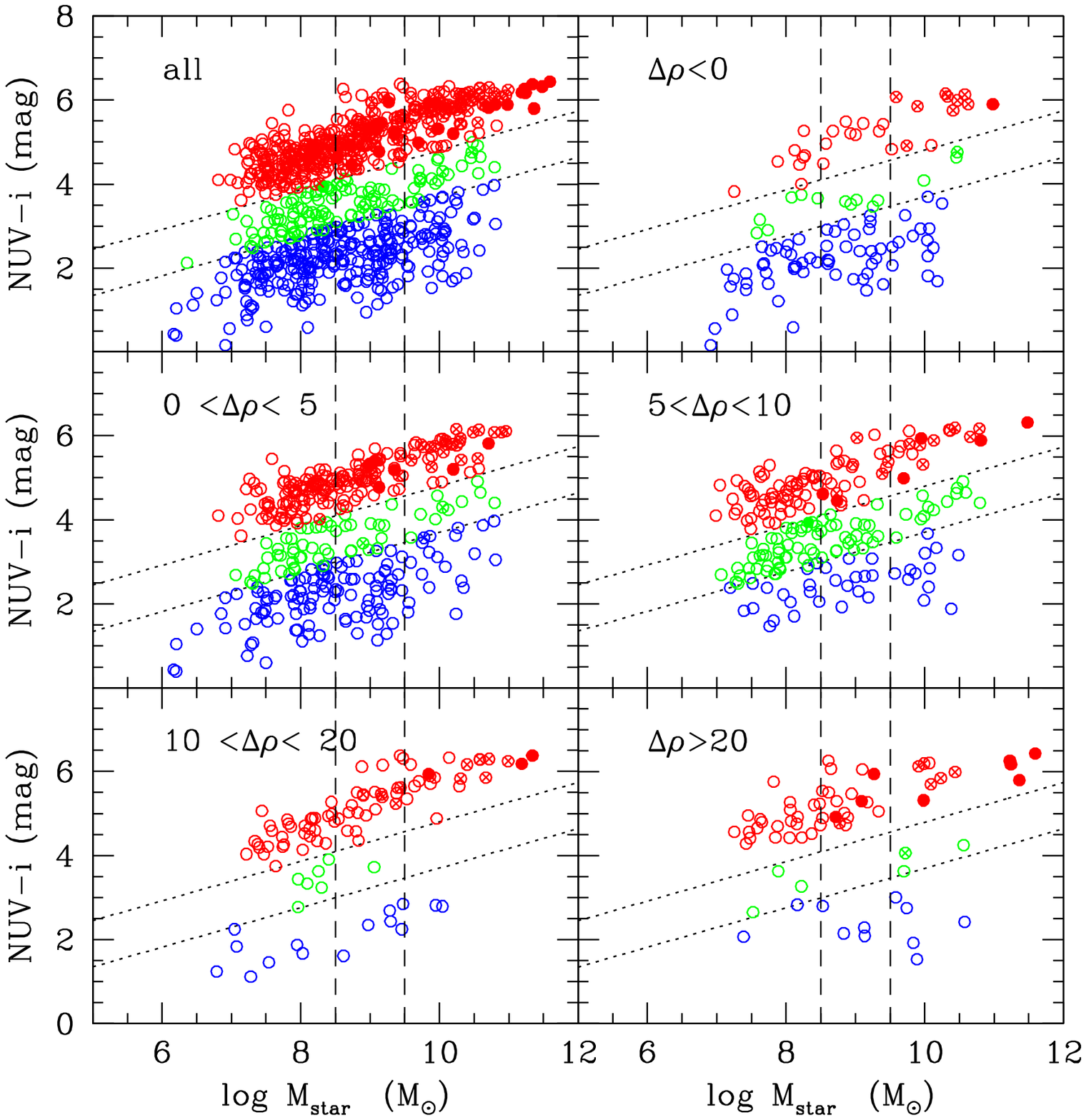}
   \caption{The extinction corrected $NUV-i$ (AB system) vs. $M_{star}$ relations for galaxies 
   selected according to their density contrast. Red, green, and blue colours are for galaxies 
   in the red sequence, green valley, and blue cloud, respectively. Filled symbols are for slow rotators, crosses for fast rotators.
   The vertical dashed lines show the interval in stellar mass used to define massive, intermediate, and low-mass objects within the sample.
   }
   \label{CMRmassdensity}%
   \end{figure*}

An alternative way of testing the dependence of the colour-stellar mass relation on the environment is by plotting it in bins of $\Delta \rho$ as first proposed
in the Coma supercluster by Gavazzi et al. (2010). This is done in Fig. \ref{CMRmassdensity}, showing 
that the membership of galaxies to the red sequence, green valley, and blue cloud depends on the density in their surrounding. The red sequence is already formed in the lowest density regions
in the outskirts of the cluster. Here, however, the fraction of red galaxies is significantly smaller than that of star-forming systems. We also notice that the number of 
massive red galaxies overcomes that of red dwarfs. 
We also observe that galaxies gradually populate the blue cloud, the green valley and, finally, the red sequence with increasing density contrast. This effect is
more pronounced in low-mass galaxies than in massive systems. In the highest density bins ($\Delta \rho$ $>$ 10), for instance, the galaxy population is dominated by red objects, 
consistently with Fig. \ref{frazione}. On the contrary, galaxies in the green valley are present mainly in the medium density bins (0 $<$ $\Delta \rho$ $<$ 10).

\subsection{The colour-stellar mass relation in the velocity space}

   \begin{figure*}
   \centering
   \includegraphics[width=18cm]{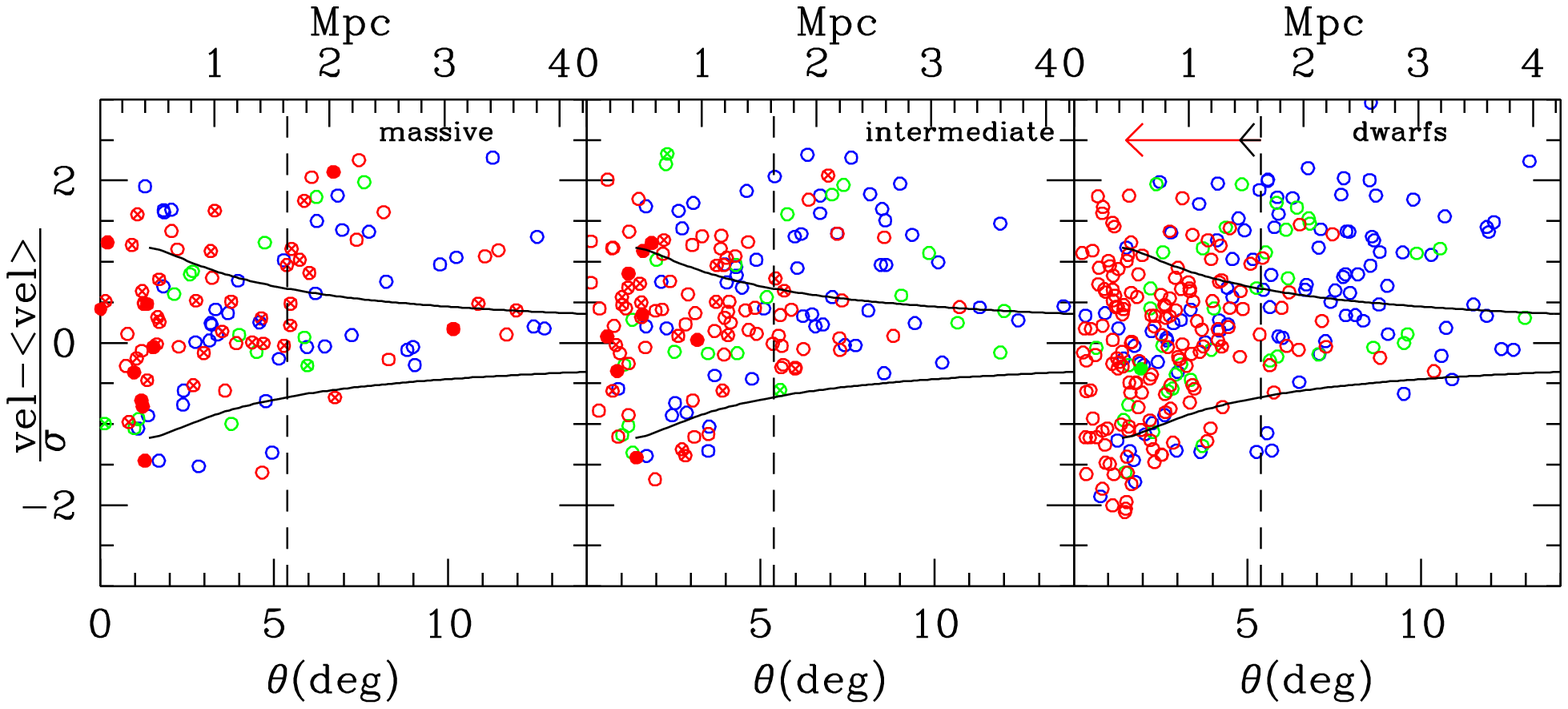}
   \caption{The relationship between $\frac{vel-<vel>}{\sigma}$ and the angular distance from the cluster centre for massive, intermediate and low-mass galaxies in cluster A and in the field.
   Red, green, and blue colours are for galaxies in the red sequence, green valley, and blue cloud, respectively. Filled symbols are for slow rotators, crosses for fast rotators. 
   The black dashed vertical line indicates the virial radius of the cluster.
   The solid lines indicate the caustics drown to identify the gravitationally bounded region of cluster A. 
   The black and red horizontal arrows show the mean distance covered by dwarf galaxies within the cluster during the time that they need to become HI-deficient (black) 
   and red (red) after the quenching of their star formation activity, as predicted by our models.}
   \label{causticamass}%
   \end{figure*}

A further step in the analysis of the spectrophotometric properties of galaxies within the cluster can be done by studying
the distribution of red sequence, green valley, and blue cloud objects in the velocity space. To do that, we plot in Fig. \ref{causticamass}
the relationship between $\frac{vel-<vel>}{\sigma}$, where $<vel>$ and $\sigma$ are the mean recessional velocity and the velocity dispersion of cluster A, 
and the angular distance from the cluster core again for galaxies belonging to cluster A
and to the field. As defined, the parameter $\frac{vel-<vel>}{\sigma}$ quantifies, for each galaxy, the excess of velocity along the line of sight
with respect to the mean velocity dispersion of the cluster, providing useful information in the third dimension, while the angular distance from the cluster centre 
gives that on the plane of the sky. To combine these two different variables, we also plot in Fig. \ref{causticamass} the caustics determined as 
indicated in van Haarlem et al. (1993):

\begin{equation}
{\frac{V_{pec}}{H_0 r} \simeq - \frac{1}{3} [\frac{<\rho_{cl}(r)>}{\rho_c}]^{0.75} \times \Omega_0^{-0.15}}
\end{equation}

\noindent
where $\rho_c$ = $\frac{3H_0^2}{8\pi G}$ is the critical density of the universe and $\rho_{cl}(r)$ the mass density distribution within the cluster.
The caustic is a useful tool to separate the infalling regions from the virialised part of the cluster.
We calculate the radial density distribution $\rho_{cl}(r)$ using the relation:

\begin{equation}
{\rho_{cl}(r) = \rho_0 [1 + \frac{r^2}{r_c^2}]^{-3\beta/2}}
\end{equation}

\noindent
assuming $r_c$ = 65 arcmin and $\beta$ = 0.75, as determined by Schindler et al. (1999) for the Virgo cluster. $\rho_0$, for which we do not have any direct estimate,
is chosen to $\sim$ match the distribution of galaxies at low $\frac{vel-<vel>}{\sigma}$ in Fig. \ref{causticamass}. 
The analysis of Fig. \ref{causticamass} gives three major results:\\
1) the inner regions of Virgo cluster A are almost totally devoid of unperturbed galaxies with a normal
star formation activity (blue cloud) and, to a lower extent, quiescent late-type 
systems belonging to the green valley, and this regardless of their stellar mass. Within the caustics, the main galaxy population is 
that of red quiescent objects. \\
2) within the central half virial radius ($\sim$ 0.8 Mpc), the galaxies with the highest peculiar
velocities with respect to the mean velocity of the cluster are red, quiescent dwarf systems. These are thus 
early-type dwarfs which recently entered the cluster. \\
3) the majority of the massive and intermediate slow rotators (filled dots) are located in the inner half virial radius and within the caustics, indicating that they are 
cluster members since the early formation of Virgo.

  \begin{figure}
   \centering
   \includegraphics[width=11cm]{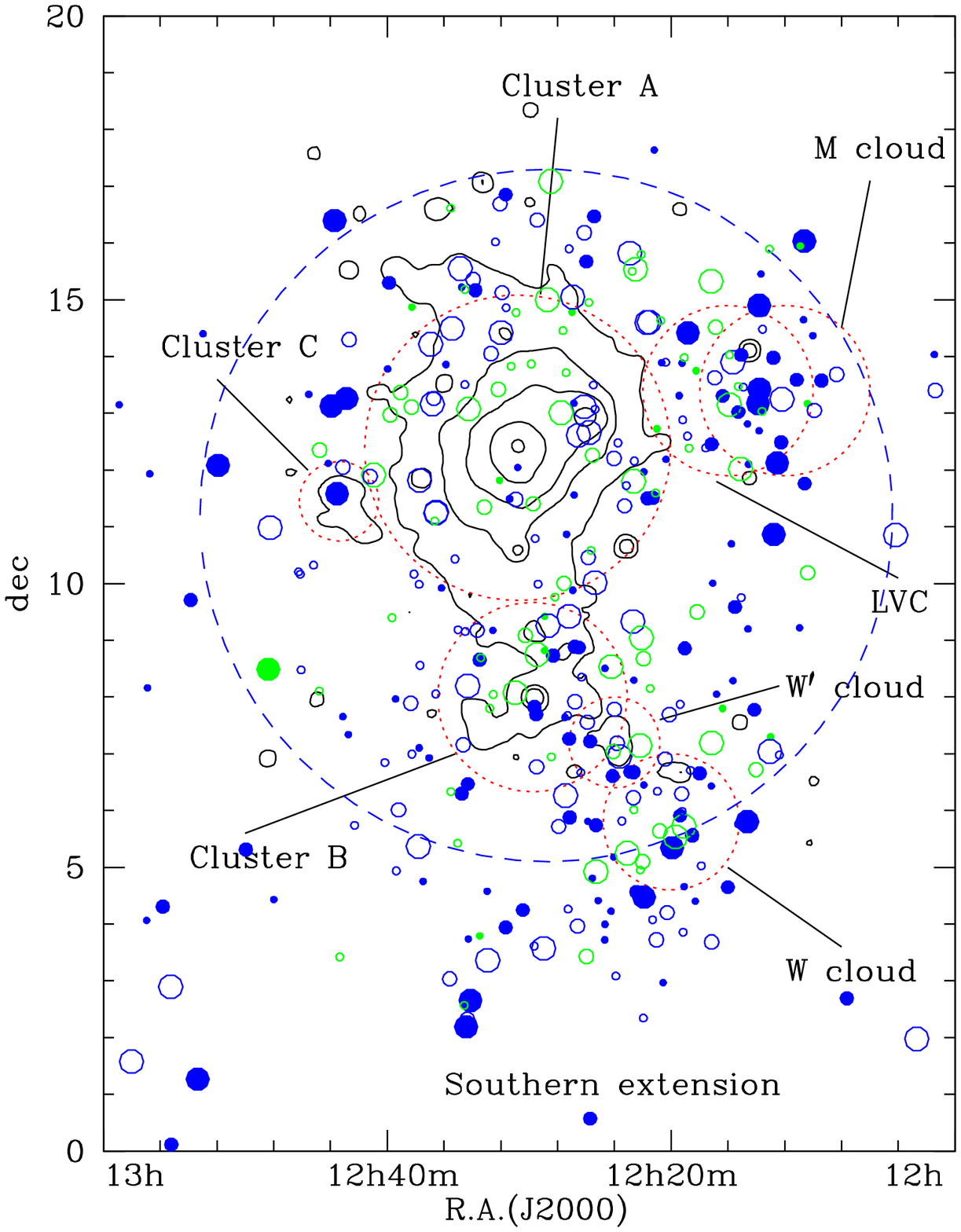}
   \caption{Sky distribution of star-forming galaxies with colours of the symbols 
   coded as in Fig. \ref{CMRmass}. Filled symbols indicate galaxies with a normal HI content ($HI-def$ $\leq$ 0.4),
   empty symbols HI-deficient galaxies ($HI-def$ $>$ 0.4).
   The size of the symbols is proportional to the stellar mass of galaxies: big symbols are for galaxies
   with $M_{star}$ $>$ 10$^{9.5}$ M$_{\odot}$, medium size symbols for objects with 10$^{8.5}$ $<$ $M_{star}$ $\leq$ 10$^{9.5}$ M$_{\odot}$, and small size symbols for 
   $M_{star}$ $\leq$ 10$^{8.5}$ M$_{\odot}$.  
    }
   \label{angdistmass3}%
   \end{figure}

\section{Discussion: a unified picture of galaxy evolution}

In this section we combine these new observational results to the prediction of our multizone chemo-spectrophotometric models of
galaxy evolution with the purpose of understanding the role of the cluster environment in the formation of the red sequence. 
The analysis presented so far suggests that massive and dwarf galaxies behave in different ways.
We thus address this discussion separately for massive and dwarf systems.

\subsection{Massive galaxies}

\begin{figure*}
   \centering
   \includegraphics[width=18cm]{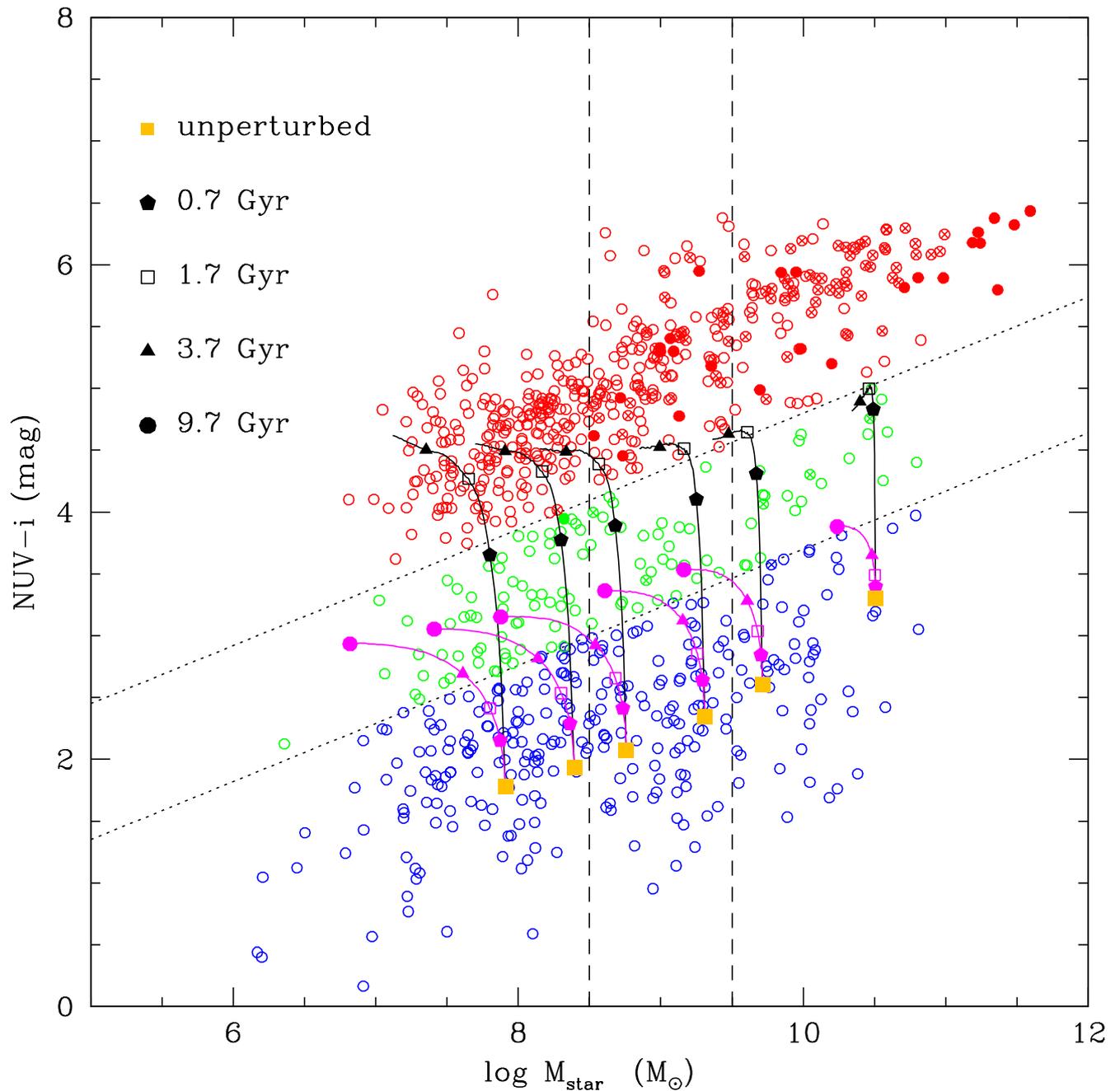}
   \caption{The extinction corrected $NUV-i$ (AB system) vs. $M_{star}$ relation for all galaxies of the sample. 
   Filled symbols are for slow rotators, crosses for fast rotators. Symbols are colour coded as in Fig. \ref{CMRmass}. The large orange filled squares indicate 
   the models of unperturbed galaxies of spin parameter $\lambda$=0.05 and rotational velocity 40, 55, 70, 100, 130, 170, and 220 km s$^{-1}$. The magenta lines 
   indicate the starvation models. The black lines show the ram pressure stripping models. 
   Different symbols along the models indicate the position of the model galaxies at a given look-back time from the beginning of the interaction. }
   \label{CMRmassmod}%
   \end{figure*}

The star formation activity of galaxies is tightly related to their content of atomic and molecular gas, as indicated by the Schmidt law (Schmidt 1959; Kennicutt 1998b; Bigiel et al. 2008).
The gaseous component, in particular the atomic phase which in spiral galaxies is located on an extended disc, can be easily removed during any kind of interaction with the hostile
cluster environment (e.g. Boselli \& Gavazzi 2006). Fig. \ref{angdistmass3} shows the distribution of HI-normal ($HI-def$ $\leq$ 0.4) and HI-deficient 
($HI-def$ $>$ 0.4) late-type galaxies within the Virgo cluster region. 
Figure \ref{angdistmass3} indicates that gas-rich, massive galaxies are not present in the high-density regions associated to the X-ray emitting gas in cluster A
(e.g. Cayatte et al. 1990). This observational result has been historically interpreted as a strong evidence of ram pressure stripping.
The same galaxies are also rare in the other high-density substructures, with the exception of the M cloud, as firstly noticed by Gavazzi et al. (1999), and in the LVC.
The lack of gas, both in its atomic (Hughes \& Cortese 2009; Cortese \& Hughes 2009; Gavazzi et al. 2013a,b) and molecular (Fumagalli et al. 2009; Boselli et al. 2014)
phase, quenches the activity of star formation, making HI-deficient cluster galaxies redder than unperturbed objects.
To see the effects of gas stripping on the stellar population of galaxies, we plot in Fig. \ref{CMRmassmod} the $NUV-i$ vs. $M_{star}$ relation for all 
galaxies of the sample and compare this diagram to the predictions of the models.
Figure \ref{CMRmassmod} shows that the ram pressure 
gas stripping due to a single crossing of the cluster is not sufficient to fully stop the activity of star formation of infalling massive late-type galaxies.
Gas deficient late-type galaxies, with a residual activity of star formation that make them fall in the green valley, are indeed still present on the innermost regions of the cluster.
The $NUV-i$ colour of these gas stripped galaxies, indeed, does not become as red as the one of similar stellar mass objects on the red sequence (Cortese \& Hughes 2009). 
As extensively discussed in Boselli et al. (2014), in these massive objects the total gas removal 
via ram pressure stripping requires timescales of the order of 1.5 Gyr, as indicated by recent hydrodynamic simulations (Tonnesen \& Brayan 2009; Roediger \& Bruggen 2007). 
These timescales have been determined using tuned simulations able to reproduce the different gas phases, from the densest molecular gas
located inside giant molecular clouds to the diffuse gas of the ISM (Tonnesen \& Bryan 2009). They also consider the change in the impact parameter that a galaxy encounters
along its orbit within the cluster (Roediger \& Bruggen 2007). These timescales are relatively short because these recent models are able to strip also the molecular gas phase,
as now confirmed by observations (Fumagalli et al. 2009; Boselli et al. 2014). Once the gas is removed, the delay in the quenching of the star formation activity
is $\sim$ 0.8 Gyr longer, thus of the order of $\sim$ 2.3 Gyr. This timescale is longer than the typical crossing time of the cluster ($\sim$ 1.7 Gyr; Boselli \& Gavazzi 2006). 
It is thus not surprising that gas-poor, freshly infalling star-forming massive galaxies are located all over the cluster since they do not have the time to totally stop their activity 
becoming red objects before reaching the cluster core. This process of migration from the blue cloud to the red sequence is now also reproduced in cosmological simulations (e.g. Cen 2014).\\

Figure \ref{CMRmassmod} also shows that galaxy starvation is not able to create the red sequence just by quenching the star formation activity through gas consumption once 
the infall of pristine gas is stopped. The models shown in Fig. \ref{CMRmassmod} produce late-type galaxies too blue compared to red sequence objects even if the 
starvation process started $\sim$ 10 Gyr ago (oldest plotted model). Using the same models, Cortese et al. (2011) have shown that starvation cannot 
reproduce the observed HI scaling relations of gas-poor cluster galaxies.
This result is fully consistent with what presented in Boselli et al. (2006) for NGC 4569. The starvation process is not able to
produce the truncated radial profiles typical of gas stripped galaxies, and is expected to significantly decrease the mean effective surface brightness of the perturbed galaxies.
Starvation is thus inconsistent with observations since it is known that the mean effective surface brightness of lenticular galaxies, the potential output of the starvation process,
is brighter than that of star-forming systems
of similar mass (Boselli \& Gavazzi 2006). In galaxies recently stripped of their gas, the timescale for gas consumption determined considering the atomic 
and molecular gas phase plus the recycled phase produced by stars during their evolution ($\tau_{gas,R}$ $\simeq$ 3.0-3.3 Gyr)
is also significantly longer than the timescale for complete gas stripping via ram pressure, making starvation a quite improbable mechanism for quenching the 
activity of star formation in massive galaxies in clusters (Boselli et al. 2014). We recall, however, that the original definition of starvation proposed by Larson et al. (1980)
considers that the consumption of gas via star formation happens once the halo of the hot gas surrounding galaxies, expected to feed the stellar disc, 
is removed during the interaction. This definition is slightly different than the one adopted in our models, which is based on the assumption that gas infall is stopped.
Our models, however, do not consider a hot halo of gas but only the gaseous component located on the disc, thus they should be quite representative of
the general definition often used in cosmological and semianalytic models.\\

\begin{figure}
   \centering
   \includegraphics[width=9cm]{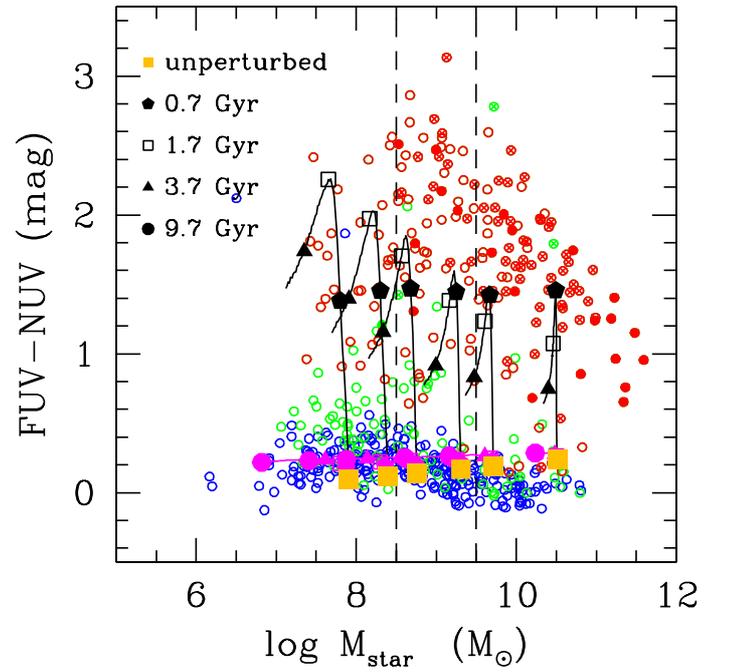}
   \caption{The extinction corrected $FUV-NUV$ (AB system) vs. $M_{star}$ relation. 
   Symbols, dotted and dashed lines, and models are as in Fig. \ref{CMRmassmod}.}
   \label{CMRFUVNUVmass}%
   \end{figure}

Figure \ref{CMRmassmod} also reveales that the most massive red galaxies, those with $M_{star}$ $\gtrsim$ 10$^{10.7}$ M$_{\odot}$, do not have similar counterparts
in the blue cloud nor in the green valley. It is thus impossible that a simple quenching of the star formation activity of a late-type galaxy forms these massive, red systems.
The quenching of the star formation activity indeed prevents the secular increase of the stellar mass typical of unperturbed objects, producing perturbed galaxies
of lower stellar mass then their isolated counterparts. The most massive red galaxies
have stellar masses as high as $M_{star}$ $\simeq$ 10$^{11.5}$ M$_{\odot}$, thus the observed shift in the bright end between the red sequence and blue cloud galaxies
cannot be due to a systematic error in the stellar mass determination of the two different populations (see Appendix B). Figure \ref{CMRmassmod} shows that the majority of these
massive ($M_{star}$ $\gtrsim$ 10$^{10.7}$) red, early-type galaxies are slow rotators. Following the arguments of Cappellari et al. (2011b) and Cappellari (2013), it is thus conceivable that these objects
have been formed through major merging events. This is a violent process that can dynamically heat the system and transform, on short timescales, the full amount of gas into stars, 
producing pressure supported, gas poor roundish objects characterised by old stellar populations (e.g Barnes 1992; Barnes \& Hernquist 1996). The $FUV-NUV$ vs. $M_{star}$ relation plotted in Fig. \ref{CMRFUVNUVmass},
which is sensitive to the presence of young populations in star-forming systems and to very old stars in evolved objects (the UV upturn, O'Connell 1999; Boselli et al. 2005)
suggests that these most massive galaxies are also those with the oldest stellar populations (Bureau et al. 2011), thus pushing the major merging event to very early epochs. 
This picture is consistent with what is shown in Fig. \ref{causticamass}, where all but one of the slow rotators of cluster A are located within the caustic,
indicating their membership to the virialised component of the cluster. They are thus within the densest regions since an early age.

\begin{figure}
   \centering
   \includegraphics[width=11cm]{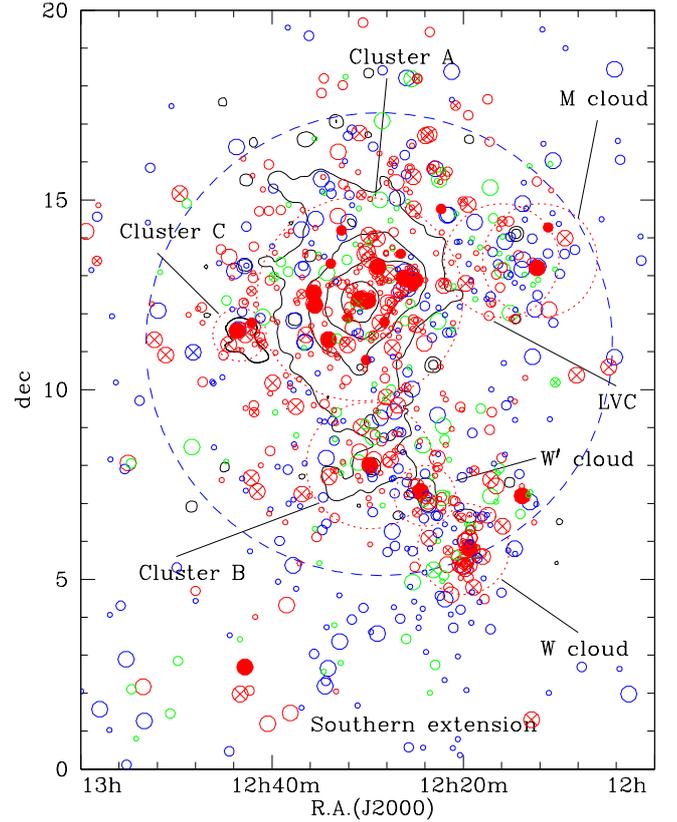}
   \caption{Sky distribution of galaxies with colours of the symbols coded as in Fig.\ref{CMRmass}. 
   Filled symbols indicate slow rotators, crosses fast rotators.
   The size of the symbols is proportional to the stellar mass of galaxies: big symbols are for galaxies
   with $M_{star}$ $>$ 10$^{9.5}$ M$_{\odot}$, medium size symbols for objects with 10$^{8.5}$ $<$ $M_{star}$ $\leq$ 10$^{9.5}$ M$_{\odot}$ and small size symbols for 
   $M_{star}$ $\leq$ 10$^{8.5}$ M$_{\odot}$.  
    }
   \label{angdistmass2}%
   \end{figure}

Figure \ref{angdistmass2} shows the distribution of galaxies within the Virgo cluster region using different symbols to identify fast and slow rotators.  
Combined with Fig. \ref{CMRmassall}, Fig. \ref{angdistmass2} shows that these slow-rotating massive early-type galaxies are the most massive objects
of each single substructure within the cluster. Indeed this is the case for M87 in cluster A, M49 in cluster B, NGC 4168 in the M cloud, NGC 4261 in the W cloud, and NGC 4365 in the W' cloud.
The most massive galaxy of Virgo cluster C, M60 (NGC 4649), is classified in Emsellem et al. (2011) as a fast rotator. We notice, however, that if we 
calculate the limiting spin parameter $\lambda_{Re}(lim)$ below which that galaxy is considered as a slow rotator, $\lambda_{Re}(lim)$ = 0.31 $\sqrt{\epsilon}$, where
$\epsilon$ is the ellipticity of the galaxy, using the photometric dataset of Cortese et al. (2012a) for the \textit{Herschel} Reference Survey in the $i$-band ($\epsilon$ = 0.19), 
$\lambda_{Re}(lim)$ = 0.135, while the observed value is $\lambda_{Re}$ = 0.127. M60 could thus be considered as a slow rotator, as all the other most massive galaxies 
of each single cluster substructure. The only exception is NGC 4216 in the LVC, which is a massive, HI-deficient spiral galaxy 
classified as SAB(s)b: in NED. We can thus conclude that, in groups characterised by velocity dispersions of 200 $\lesssim$ $\sigma$ $\lesssim$ 550 km s$^{-1}$, 
all the most massive galaxies ($M_{star}$ $\gtrsim$ 10$^{11}$ M$_{\odot}$) have been formed by major mergers. Among these groups, those included 
in the \textit{ROSAT} image of B\"ohringer et al. (1994) are also charaterised by a diffuse X-ray emission. Being the X-ray gas a direct tracer of the potential well,
this is the first direct evidence that the dominant galaxies of massive haloes are slow rotators, as indeed expected but never observed so far (Scott et al. 2014). 
This picture is consistent with the semianalytic m§odels
of De Lucia et al. (2006), which indicate that only the most massive early-type galaxies ($M_{star}$ $\gtrsim$ 10$^{11}$ M$_{\odot}$) are formed by major merging events.
It is also consistent with the semianalytic predictions of Khochfar et al. (2011), which indicate that slow rotators 
are preferentially central galaxies associated with a large halo.
It also matches with the observational work of Wilman \& Erwin (2012), who concluded using different arguments that central galaxies of massive haloes are ellipticals formed by major merging events.
On a more general context, they also confirm that slow rotators are primarily situated in high-density regions (Cappellari et al. 2011, Cappellari 2013, Houghton et al. 2013, 
D'Eugenio et al. 2013, Scott et al. 2014).

In the Virgo cluster A, however, there are quite a few galaxies with stellar masses down to $\sim$ 10$^{10}$ M$_{\odot}$ ($\sim$ the limiting stellar mass of the
ATLAS$^{3D}$ survey) that are slow rotators, and at the same time there are 
quite a few fast rotators with $M_{star}$ as high as $\sim$ 10$^{11}$ M$_{\odot}$. Slow rotators are mainly located inside the inner half virial radius (Fig. \ref{angdistmass2}), and are 
preferentially virialised within the cluster potential (Fig. \ref{causticamass}). It is thus conceivable that they have been formed by a merging event early in the past, when galaxies and 
groups where smaller than at the present epoch. The most massive fast rotators have stellar masses not significantly larger than those of star-forming discs, and thus we cannot exclude
that they have been formed through less violent phenomena. The considerations given above on the higher effective surface brightness of massive lenticulars with respect to that of
similar late-type systems suggest that gravitational interactions, efficient at early epochs when galaxies were accreted in infalling groups (pre-processing), 
rather than a simple ram pressure stripping event, is at the origin of this population 
(e.g. Boselli \& Gavazzi 2006; Dressler 2004; Wilman et al. 2009; Boundy et al. 2010; Dressler et al. 2013).
Multiple gravitational interactions (galaxy harassment, Moore et al. 1998) are indeed able to induce the formation of bars, 
with nuclear gas infall, thus producing thicker discs than those formed through a mild, secular evolution.
Clearly a full understanding of the kinematical properties of the Virgo cluster galaxies would benefit from a larger dataset
also including late-type systems.

\subsection{Dwarf galaxies}

This new set of observations and the models are fully consistent with the evolutionary picture originally presented in Boselli et al. (2008a,b) for dwarf systems. 
Boselli et al. (2008a) and Gavazzi et al. (2013a) estimated a high infall rate of dwarf galaxies in the Virgo cluster ($\sim$ 300 galaxies Gyr$^{-1}$).
The ram pressure stripping exerted by the hot 
and dense intergalactic medium on these dwarf galaxies moving at high velocity within the cluster is sufficient to fully remove, on very short timescales ($\lesssim$ 200 Myr), 
their total gaseous component, as indeed shown in Fig. \ref{angdistmass3}. 
Because of the lack of gas, the infalling galaxies stop their activity of star formation, becoming red systems in $\simeq$ 1 Gyr. 
Figure \ref{CMRmassmod} clearly shows that this process is so efficient to transform star-forming galaxies in quiescent systems as red as those populating the red sequence.
The efficiency even increases with multiple crossings of the cluster (Boselli et al. 2008a). The analysis presented so far, made possible thanks to the full coverage of the Virgo cluster region
at different frequencies, confirms this scenario. The most recent hydrodynamical simulations (Roediger \& Bruggen 2007, Tonnesen \& Bryan 2009; Cen et al. 2014)
and the observation of star-forming galaxies with HI (e.g. Scott et al. 2012), H$\alpha$ (e.g. Yagi et al. 2008), and X-ray (Sun et al. 2010) tails of stripped material at the periphery of 
clusters consistently indicate that ram pressure stripping is efficient even outside the virial radius of massive clusters such as Virgo. Given that this process
is able to transform star-forming systems into red objects on timescales $\simeq$ 1 Gyr, and that these freshly infalling objects are moving on the plane of the sky with a velocity with respect to 
the cluster centre of $\sigma$=1150 km s$^{-1}$ (the typical velocity dispersion of late-type galaxies in the cluster, from Boselli \& Gavazzi 2006), we expect that most of them 
are fully transformed before they reach the cluster core. This would thus explain the observed variation of the red-to-blue fraction with the angular distance from the cluster centre or
the trend with local density observed in Fig. \ref{frazione} and Fig. \ref{CMRmassdensity}. This picture is also consistent with what observed in Fig. \ref{causticamass}.
The presence of red dwarf galaxies within the inner region of cluster A but outside the caustic, thus not belonging to 
the virialised component, suggests that the infall of these systems is relatively recent.\\

\begin{figure*}
   \centering
   \includegraphics[width=0.33\textwidth]{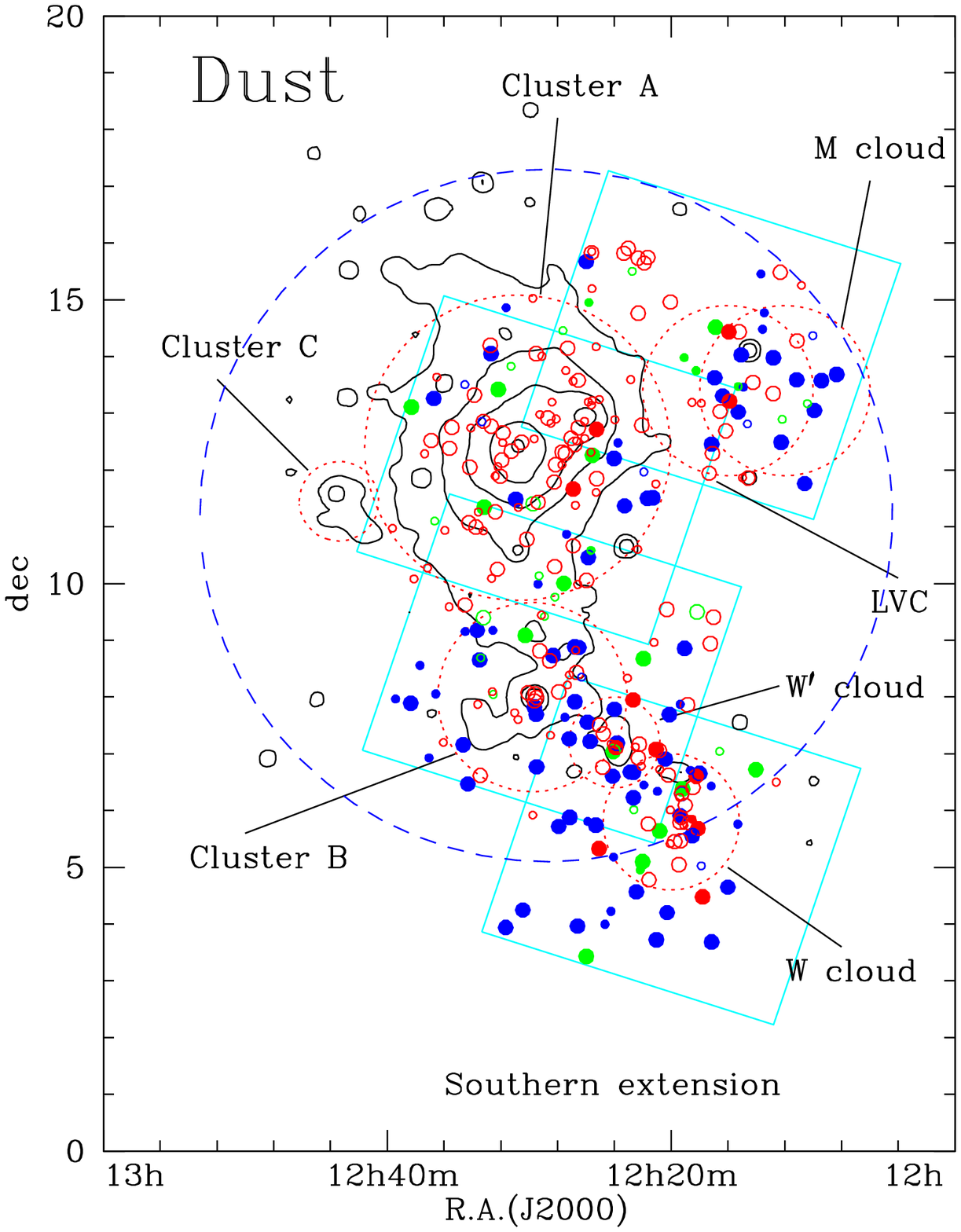}
   \includegraphics[width=0.33\textwidth]{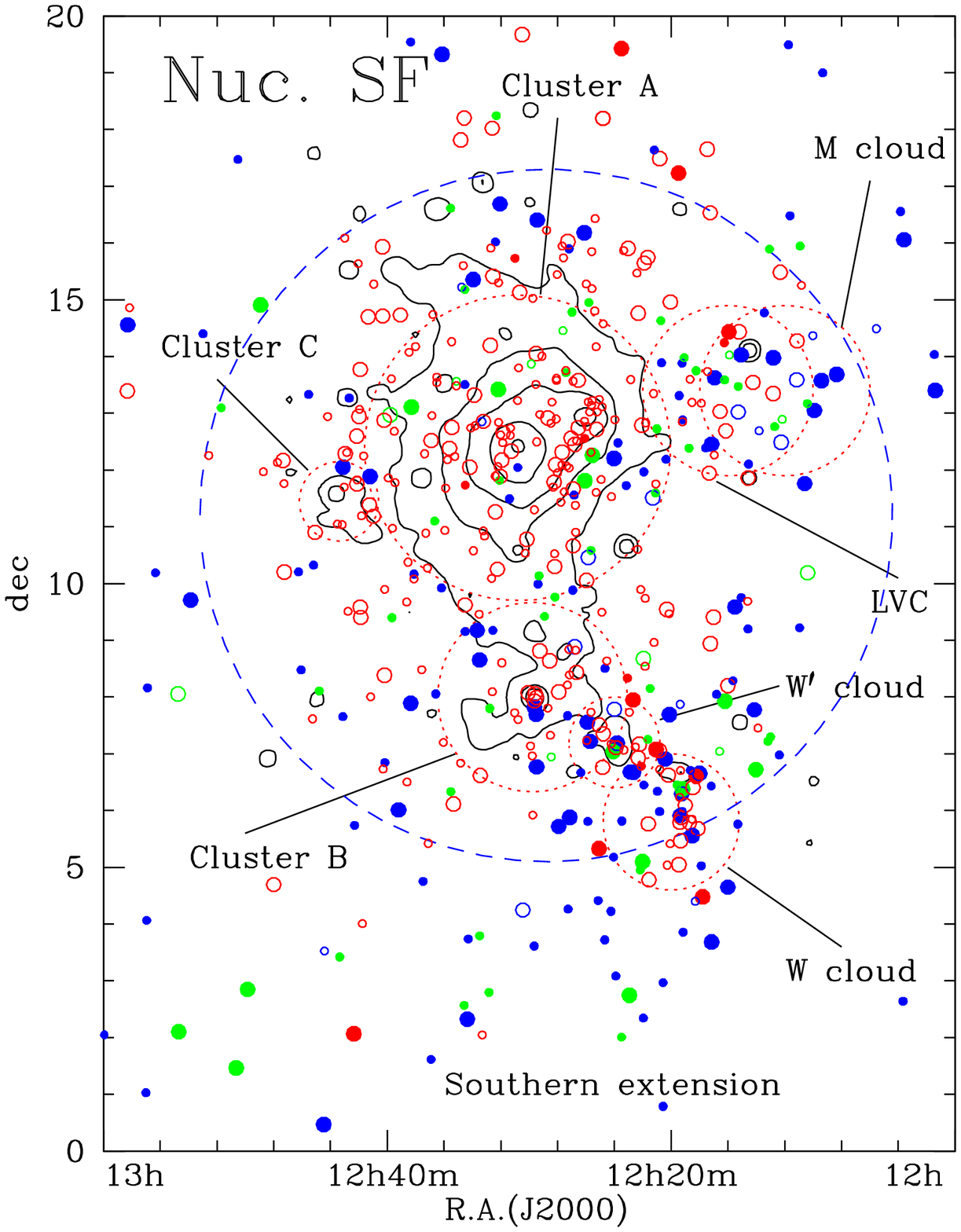}
   \includegraphics[width=0.33\textwidth]{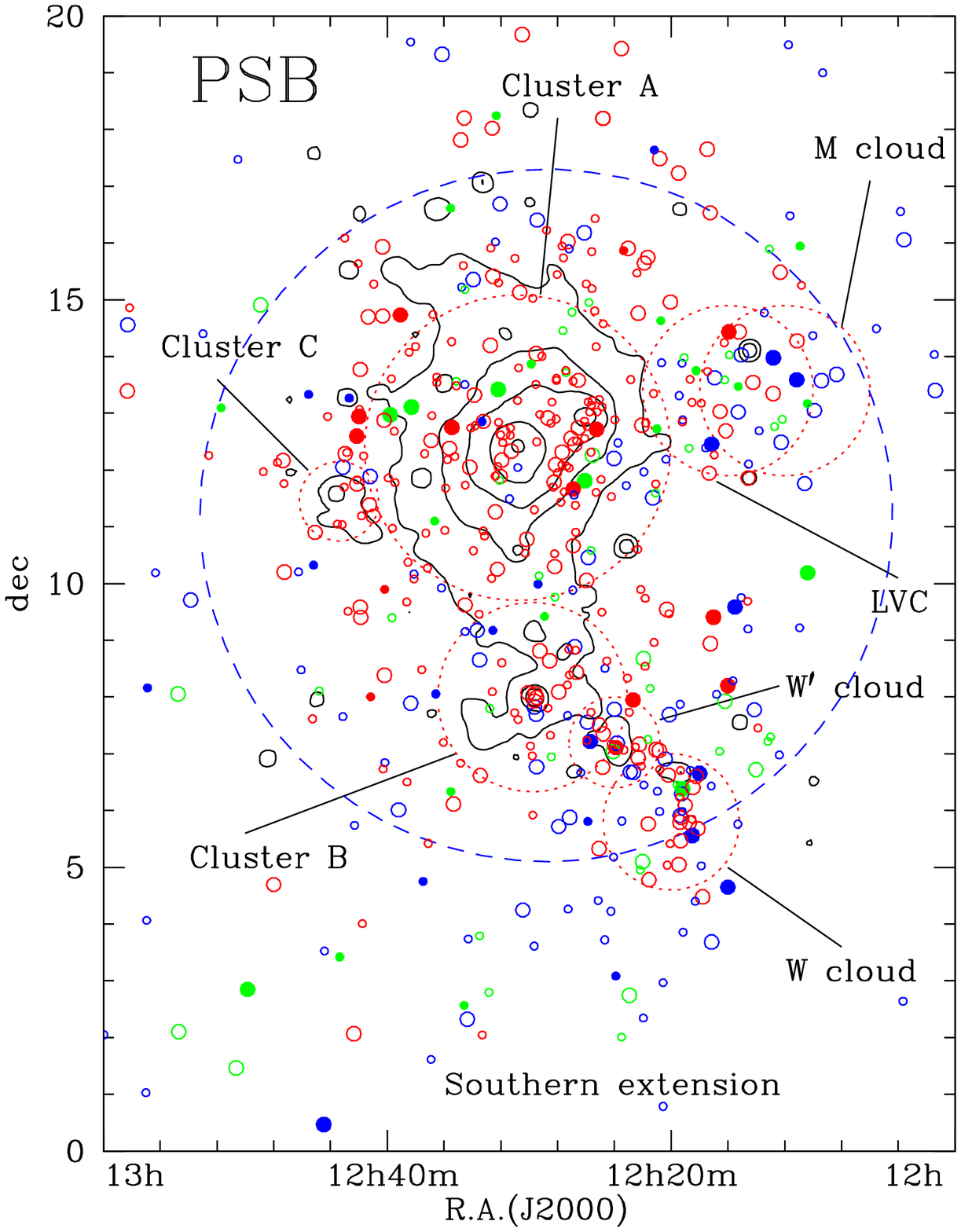}\\
   \caption{Left: sky distribution of the subsample of galaxies observed by \textit{Herschel}. Red, green, and blue symbols are used
   to indicate galaxies belonging to the red sequence, green valley, and blue cloud, respectively. Filled dots indicate dust-rich galaxies, while empty symbols undetected objects.
   The size of the symbols is proportional to the stellar mass of galaxies: medium size symbols are for objects with 10$^{8.5}$ $<$ $M_{star}$ $\leq$ 10$^{9.5}$ M$_{\odot}$,
   small size symbols for 10$^{8}$ $\leq$ $M_{star}$ $\leq$ 10$^{8.5}$ M$_{\odot}$. 
   Centre: sky distribution of the subsample of galaxies observed in spectroscopic mode by the SDSS. Filled dots indicate galaxies with an equivalent width in the 
   H$\alpha$ emission line $E.W.H\alpha_{emi}$ $>$ 3 \AA ~(and S/N $>$ 5), witnessing a nuclear star-forming activity. 
   Right: same as central plot but where filled dots indicate PSB galaxies.
   In these last two plots medium size symbols are for objects with 10$^{8.5}$ $<$ $M_{star}$ $\leq$ 10$^{9.5}$ M$_{\odot}$,
   small size symbols for $M_{star}$ $\leq$ 10$^{8.5}$ M$_{\odot}$.   
   }
   \label{geografia}%
   \end{figure*}

\begin{figure}
   \centering
   \includegraphics[width=9cm]{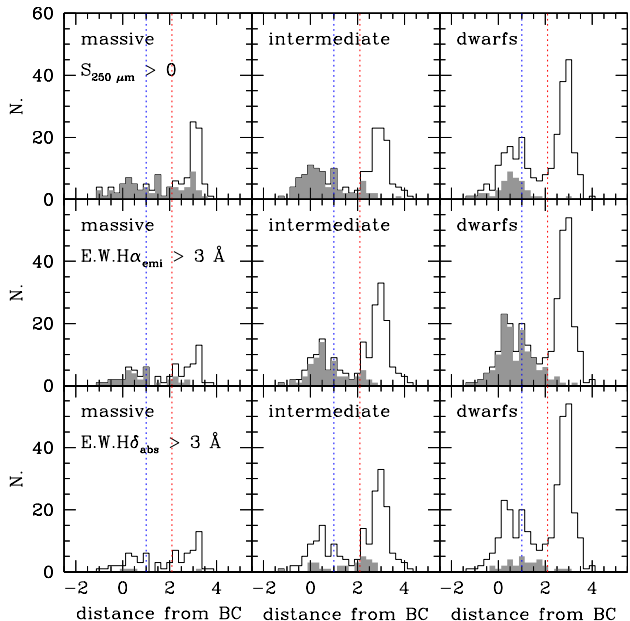}
   \caption{Distribution of the colour difference $(NUV-i)$ - $(NUV-i)_{Mod}$ corresponding to the mean distance from typical colour-stellar mass relation
   of blue cloud (BC) objects for massive ($M_{star}$ $>$ 10$^{9.5}$ M$_{\odot}$; left), intermediate (10$^{8.5}$ $<$ $M_{star}$ $\leq$ 10$^{9.5}$ M$_{\odot}$; centre), 
   and dwarf ($M_{star}$ $\leq$ 10$^{8.5}$ M$_{\odot}$; right) galaxies. The upper row shows the distribution of galaxies observed (black histogram) and detected (grey shaded histogram)
   by \textit{Herschel}, the middle and lower rows the distribution of objects with available nuclear spectroscopy from SDSS (black) with a nuclear star formation activity (grey) 
   or in a post-starburst phase (grey), respectively.
   The vertical red and blue dotted lines indicate the limits used to identify the red sequence and the blue cloud.
   }
   \label{histo}%
   \end{figure}

The unique coverage at different wavelengths of the whole Virgo cluster region can be used to test whether this scenario is consistent also with other observations. To do that we plot in Fig. \ref{geografia}
the distribution of dwarf and intermediate stellar mass galaxies ($M_{star}$ $\leq$ 10$^{9.5}$ M$_{\odot}$) within the Virgo cluster region using different symbols to indicate
whether they contain some cold dust in their interstellar medium, they have a residual nuclear star formation activity or they are in a post-starburst phase (see sect. 3). 
Dusty galaxies are identified using the 250 $\mu$m SPIRE band emission (detected galaxies), where the sensitivity of \textit{Herschel} is at its 
maximum\footnote{For this particular band we limit the analysis to those dwarf galaxies potentially detectable by 
\textit{Herschel} (10$^{8}$ $\leq$ $M_{star}$ $\leq$ 10$^{9.5}$ M$_{\odot}$).}.
We also plot in Fig. \ref{histo} the distribution of the colour difference $(NUV-i)$ - $(NUV-i)_{Mod}$, where $(NUV-i)_{Mod}$
is the colour of the unperturbed model galaxy (see sect. 5), for the galaxies observed by \textit{Herschel} (upper panels), with nuclear star formation activity (central panels),
and for post-starbursts (PSB; lower panels), where galaxies are separated in three bins of stellar mass. We recall that this plot shows the mean displacement in $NUV-i$ 
colour from the typical blue cloud colour-stellar mass relation. Given the homogeneous depth of the HeViCS survey,
the \textit{Herschel} data can be used to identify dust-rich galaxies over the mapped region, which unfortunately covers only a fraction of the GUViCS fields (see Fig. \ref{angdistmass1}). 
Figure \ref{geografia} shows that dusty galaxies, in particular those of low stellar mass, tend to avoid the X-ray emitting regions within cluster A and B. 
We notice, however, several dust-rich dwarf red galaxies in those cluster substructures not characterised by a diffuse X-ray emission (W, M, and LV clouds).  
Figure \ref{histo} indicates the presence of dust in spiral galaxies in the blue cloud and in the green valley down to $\sim$ 10$^8$ M$_{\odot}$, 
while most of the detected red galaxies have stellar masses larger than $\sim$ 10$^{9.5}$ M$_{\odot}$ (De Looze et al. 2013). The fraction of dusty galaxies
with red colours, however, continuously decreases with decreasing stellar mass. In the intermediate stellar mass range (10$^{8.5}$ $\lesssim$ $M_{star}$ $\lesssim$ 10$^{9.5}$ M$_{\odot}$),
the few objects detected at 250 $\mu$m belonging to the red sequence are those with the bluest colours.
In the lower stellar mass range ($M_{star}$ $\lesssim$ 10$^{8.5}$ M$_{\odot}$) none of the red sequence objects have been detected by \textit{Herschel}, while only half of those
belonging to the green valley and the majority in the blue sequence. 

Figure \ref{geografia} shows also the presence of several red galaxies with a residual nuclear 
star formation activity or in a post-starburst phase. They are mainly located in the cluster periphery or in those substructures not characterised by a diffuse X-ray emission 
(W, M, and LV clouds). They often corresponds to the dust-rich red galaxies detected by \textit{Herschel}.
Within cluster A (at a radial distance smaller than half the virial radius, roughly corresponding to the X-ray emitting region), 
the number of dwarf galaxies of stellar mass 10$^8$ $<$ $M_{star}$ $<$ 10$^{9.5}$ M$_{\odot}$  observed by \textit{Herschel} down to its detection limit with spectroscopic data is 78.
Out of these, only 11 are dust-rich objects (14\%). Among those with a nuclear star formation activity, the fraction of dusty objects 
increases up to 58\% (7/12), while among PSB dwarf far-infrared detected objects are 62\% (5/8). Similar ratios are obtained when the analysed region is extended to the whole
\textit{Herschel} field, thus adding the other main substructures of the cluster (Fig. \ref{geografia}). Figure \ref{histo} shows that PSB galaxies, thus those objects that recently 
abruptly stopped their nuclear star formation activity, are galaxies of intermediate or low stellar mass. Although present in the blue cloud, PSB galaxies are preferentially
located in the green valley or are among the bluest objects in the red sequence (Gavazzi et al. 2010). They are frequent only at low and intermediate stellar masses. Galaxies with a nuclear star 
formation activity are preferentially late-type systems both belonging to the blue cloud and to the green valley. This indicates that the quenching of the star formation activity 
is principally in the outer disc of spirals. There are, however, quite a few dwarf early-type galaxies in the red sequence of intermediate or low stellar mass 
where star formation is still present in their nucleus.

To summarise, the observations collected so far combined with hydrodynamical simulations and our specrophotemetric models of galaxy evoultion, consistently indicate that
the ram pressure exerted on the dwarf galaxy population infalling into the cluster is able to remove the atomic gas on short timescales. The stripping process
start to be efficient outside the virial radius, as indicated by the presence of HI-deficient galaxies up to $\sim$ 2 virial radii (Gavazzi et al. 2013a and Fig. \ref{angdistmass3}).
In a ram pressure stripping process both gas and dust are stripped outside-in (e.g. Boselli et al. 2006; Fumagalli \& Gavazzi 2008; Cortese et al. 2010, 2012b).
Some gas and dust can be retained in the nuclear regions, where the potential well is at its maximum, as indicated by high resolution \textit{Herschel} images (de Looze et al. 2010; 2013). 
This happens mainly in the most massive dwarfs or in those objects located outside the X-ray emitting regions, where gas stripping is expected to be less efficient. Gas removal quenches the activity 
of star formation, making galaxies redder. The activity of star formation, however, can last longer in the nuclear regions where some gas and associated dust can be still retained.

This scenario is also confirmed by the recent kinematic observations of Toloba et al. (2009, 2011, 2012, 2014). Indeed a large fraction of dwarf elliptical galaxies 
with stellar masses 10$^{8.5}$ $<$ $M_{star}$ $<$ 10$^{9.5}$ M$_{\odot}$ within Virgo are rotationally supported systems with rotation curves similar 
to those of late-type systems of comparable stellar mass. 
In a soft, ram pressure stripping event the angular momentum must be conserved, thus without other violent gravitational interactions with
nearby companions the perturbed galaxies should conserve their rotation curve. In this picture, fast rotators have been formed after a mild ram pressure stripping event 
able to remove the whole gas content of a low-mass rotating disc, while slow rotators are the results of more violent gravitational interactions. 
As firstly noticed by Toloba et al. (2009), fast rotators are preferentially situated at the periphery of the 
cluster. Contrary, slow rotators are preferentially located close to the highest density regions, within the X-ray emitting gas in cluster A 
(8 out of the 10 slow rotators of the sample; see Fig. \ref{angdistmass2}). They are also mainly situated within half the virial radius and within the caustics (7/10), suggesting that
they belong to the virialised population inhabiting the cluster since its formation (Conselice et al. 2001). It is thus conceivable that the harassment induced by multiple encounters with other members 
and with the potential of the cluster as a whole, which is efficient on relatively long timescales, had the time to perturb these galaxies, producing 
roundish, high surface brightness, pressure supported objects (Benson et al. 2014; Toloba et al. 2014). Harassment, however, can hardly explain the formation of the whole dE galaxy population of Virgo.
Indeed, the structural properties of harassed galaxies, as predicted by the models of Mastropietro et al. (2005), do not match the observed properties of typical dE in Virgo, instead well
reproduced in a ram pressure stripping scenario (Boselli et al. 2008b). Furthermore, the long timescales required to perturb galaxies with multiple flyby encounters does not match 
with a recent formation of the faint end of the red sequence as determined from the observations of high- and intermediate-redshift clusters 
(Kodama et al. 2004; De Lucia et al. 2004, 2007, 2009; Stott et al. 2007, 2009; Gilbank \& Balogh 2008; Jaffe et al. 2011), nor with the large infall rate of star-forming low-mass systems inferred by Boselli et al. (2008a)
and Gavazzi et al. (2013a). There is also additional evidence against pre-processing through galaxy harassment in dwarf galaxies within the groups infalling onto Virgo. Indeed, in these groups
the red-to-blue galaxy ratio is significantly smaller than the one observed in the core of Virgo A, where most of the dE are located (see Fig. \ref{frazione}), suggesting that
harassment is not particularly efficient in stopping the activity of star formation within the infalling groups. This is probably related to the fact that, within these groups, both the density
and the velocity dispersion are significantly smaller than in massive clusters, thus keeping the process efficient only on long timescales.
On the other hand, fast rotators are preferentially located outside half the 
virial radius and the caustic (13/21), suggesting that they are rotating systems that have recently fallen into the cluster. All their structural, spectrophotometric and kinematic properties
can be easily explained by a mild but rapid transformation after a ram pressure stripping event.

\section{Conclusion}

We study the origin of the red sequence in high-density environments using a multifrequency analysis of a large sample of galaxies located in the Virgo cluster and in its surroundings.
The sample, which includes 868 galaxies extracted from the GUViCS survey (Boselli et al. 2011), is composed of galaxies spanning a wide range in stellar mass (10$^7$ $\lesssim$ $M_{star}$ $\lesssim$ 10$^{11.5}$ M$_{\odot}$)
and morphological type (from dwarf irregulars to massive ellipticals). We first identify the different substructures composing the cluster: Virgo cluster A, B and C, the W, W$'$, M and the low velocity clouds (LVC).
We identify galaxies belonging to the red sequence, the green valley, and the blue cloud using their dust attenuation corrected $NUV-i$ colour index. We then study how 
galaxies belonging to these three different populations are located within the various cluster substructures. We observe a clear colour-segregation effect, with red galaxies located principally
in the highest density regions (cluster A, B, C, W, and W$'$), while blue star-forming systems mainly in the periphery of the cluster.

The most massive galaxies of the sample ($M_{star}$ $\gtrsim$ 10$^{11}$ M$_{\odot}$) are all early-type galaxies dominating the different substructures of the cluster, with exception of the M and LVC clouds.
The majority of these substructures are also characterised by a diffuse emission of the intracluster gas, indicating that the associated massive ellipticals are also the dominant galaxies 
within the group dark matter halo.
These galaxies are all pressure supported systems (slow rotators) probably formed by a major merging event occurred at early epochs, as suggested by their very old stellar populations 
at the origin of their pronounced UV-upturn. The other slow rotators of the sample are galaxies of intermediate-to-high stellar mass (10$^{8.5}$ $\lesssim$ $M_{star}$ $\lesssim$ 10$^{11}$ M$_{\odot}$)
generally located within the core of the different cluster substructures, but mainly in cluster A. Their recessional velocity distribution and position indicate that they are 
virialised systems within the cluster potential well, and are thus cluster members since early epochs.

All galaxies located within the inner regions of cluster A, where the density of the intracluster medium is at its maximum, are devoid of their gaseous and dusty phases of the ISM.
Our models of galaxy evolution tailored to predict the effects of the cluster environment indicate that the ram pressure exerted by the dense ICM on galaxies moving at high velocity 
within the cluster is able to remove their gaseous component of the ISM. The lack of gas induces a mild and gradual transformation of star-forming systems 
that have recently entered the cluster into quiescent red objects. The process is particularly efficient in dwarf systems, that are easily perturbed because of their shallow potential well.
Here the migration from the blue cloud to the red sequence is very rapid ($\sim$ 1 Gyr).
This timescale is comparable to the time needed by a galaxy moving at $\sim$ 1000 km s$^{-1}$, the typical velocity dispersion of the cluster, to travel from 
the periphery to the cluster core and thus explain the observed steep increase of the red-to-blue dwarf galaxy fraction observed in the core of Virgo. 
This mild process does not perturb the kinematic properties of the transformed galaxies, that 
indeed conserve rotation curves similar to those of their star-forming analogues (Toloba et al. 2011; 2014). The distribution of dwarf red galaxies in the velocity space indicates that they
recently entered the cluster. These freshly transformed objects can thus be at the origin of the faint end of the red sequence. 
 
We can thus conclude that the most massive early-type galaxies in Virgo ($M_{star}$ $\gtrsim$ 10$^{11}$ M$_{\odot}$) are formed through major merging events that occurred far in the past. 
The fraction of objects in the red sequence formed through a major merging event, identified thanks to their kinematic properties (slow rotators), rapidly decreases with decreasing stellar mass.
These objects were probably formed during the assembly of the cluster through the merging of smaller groups (pre-processing), or are the result of galaxy harassment that
occurred in galaxies belonging to the cluster since its earliest phases. With decreasing stellar mass, the fraction 
of red galaxies formed after a ram pressure stripping event able to remove the gaseous component of late-type galaxies and quench their activity becomes dominant.

Our analysis also shows that starvation, modeled by stopping gas infall in cluster galaxies, has much smaller effects
even assuming that the process has started $\sim$ 10 Gyr ago. This process thus cannot explain the quenching of the star formation activity observed in Virgo cluster galaxies,
and thus is hardly at the origin of the red sequence.

\begin{acknowledgements}

This research has been financed by the French ANR grant VIRAGE.
We wish to thank the GALEX Time Allocation Committee for the generous allocation of time devoted to this project.
We thank teh referee for constructive comments.
MF acknowledges support by the Science and Technology Facilities Council [grant number ST/L00075X/1].
This research has made use of the NASA/IPAC Extragalactic Database (NED) 
which is operated by the Jet Propulsion Laboratory, California Institute of 
Technology, under contract with the National Aeronautics and Space Administration
and of the GOLDMine database (http://goldmine.mib.infn.it/).
GALEX (Galaxy Evolution Explorer) is a NASA Small Explorer, launched in 2003 April. We gratefully acknowledge NASA support for construction, operation, and 
science analysis for the GALEX mission, developed in cooperation with the Centre National d'Etudes Spatiales of France and the Korean Ministry of Science and Technology.
Funding for the SDSS and SDSS-II has been provided by the Alfred P. Sloan Foundation, the Participating Institutions, the National
Science Foundation, the U.S. Department of Energy, the National Aeronautics and Space Administration, the Japanese Monbukagakusho, the
Max Planck Society, and the Higher Education Funding Council for England. The SDSS Web Site is http://www.sdss.org/.
The SDSS is managed by the Astrophysical Research Consortium for the Participating Institutions. The Participating Institutions are the
American Museum of Natural History, Astrophysical Institute Potsdam, University of Basel, University of Cambridge, Case Western Reserve
University, University of Chicago, Drexel University, Fermilab, the Institute for Advanced Study, the Japan Participation Group, Johns
Hopkins University, the Joint Institute for Nuclear Astrophysics, the Kavli Institute for Particle Astrophysics and Cosmology, the Korean
Scientist Group, the Chinese Academy of Sciences (LAMOST), Los Alamos National Laboratory, the Max-Planck-Institute for Astronomy (MPIA), the
Max-Planck-Institute for Astrophysics (MPA), New Mexico State University, Ohio State University, University of Pittsburgh,
University of Portsmouth, Princeton University, the United States Naval Observatory, and the University of Washington.
RG and MPH are supported by NSF grant AST-0607007 and by a grant from the Brinson Foundation.
LC acknowledges support under the Australian Research Council's Discovery Projects funding scheme (project number 130100664).

\end{acknowledgements}

\begin{appendix}
\section{The WISE data}


The NASA's Wide-field Infrared Survey Explorer (\textit{WISE}, Wright et al. 2010) is a 40 cm space telescope with a field of view of 47'.
\textit{WISE} made an all sky survey in four photometric bands in the near and mid-infrared domain
(3.6, 4.5, 12, 22 $\mu$m). The survey has been done by mapping the sky with 8.8 seconds exposures, with on average twelve exposures
per position, which allowed to reach a point source sensitivity better than 4 mJy (5$\sigma$) at the ecliptic, significantly better at the ecliptic poles because of the longer exposures.
The angular resolution at 22 $\mu$m is 12 arcsec, while the pixel size of the co-added images is 1.375".
Recently completed, the survey provides the community with fully reduced images that can be used to extract fluxes for all kinds of sources.
Flux extraction within appropriate apertures is indeed required for the galaxies analysed in this work because of their extended nature. Flux densities provided by the 
\textit{WISE} pipeline generally underestimate their flux densities since the pipeline is optimised for point-like sources.
For the purpose of this work we retrieved from the \textit{WISE} Science Archive\footnote{http://wise2.ipac.caltech.edu/docs/release/allsky/} the images of the whole Virgo cluster region
in the 22 $\mu$m $W4$ band. This band is close to the 25$\mu$m \textit{IRAS} and 24 $\mu$m \textit{Spitzer} bands, and can be easily used, after simple corrections, 
to quantify the attenuation in the FUV and NUV bands of \textit{GALEX} using the prescription of Hao et al. (2011).\\

Fluxes have been extracted using exactly the same procedure adopted in Voyer et al. (2014) for the UV fluxes. Aperture photometry, indicated for extended sources 
such as those analysed in this work, has been done using the DS9/Funtools program \textit{Funcnts}. This tool requires the use of different apertures, one centered on the emitting source
encompassing the total emission, the others on the surrounding regions to estimate the sky contribution to the emission. For this purpose we used exactly the same apertures, 
both on source and on the sky, determined on the UV images by Voyer et al. (2014). These apertures were manually defined to include the total galaxy emission (elliptical aperture)
and to avoid in the sky other emitting sources. We choose to use the same apertures as those used in the UV bands for several reasons.
First of all, to avoid aperture effects in the extinction correction, the UV and infrared emitting flux must come from the same emitting regions. The shape  
of the on source elliptical aperture (size, orientation) has been determined on the UV images of these extended sources to fully include the galaxy emission. 
Given the similar mid-infrared and UV morphology of galaxies, the same aperture is well adapted also in the 22 $\mu$m band. The on source aperture is sufficiently large to encompass
the total mid-infrared emission which might be slightly more extended than the UV emission because of the higher resolution of \textit{GALEX} ($\sim$ 5 arcsec).
The sky regions have been selected on the UV images to avoid contaminating sources such as background galaxies or nearby companions, as well as stars, quite rare in the
UV bands at high Galactic latitudes. With the exception of stars, whose contribution in the \textit{WISE} spectral domain under study is negligible, 
the nature of the possible emitting sources in the 22 $\mu$m band is similar to that of the UV sources. Both UV and mid-infrared images can be also contaminated by a low surface brightness,
diffuse emission of Galactic cirri. The position of the sky regions around the target galaxy allows an accurate determination of the sky emission, thus to minimise any systematic
effect related to this diffuse component.\\

\begin{table}
\caption{Cumulative and differential \textit{WISE} detection rate.}
\label{WISEstat}
{
\[
\begin{tabular}{cccc}
\hline
\noalign{\smallskip}
\hline
Cumulative			&	&	&	\\
\hline
$M_{star}$ range		& All	& E-S0a	& Sa-Im-BCD \\
M$_{\odot}$			& \%	& \%	& \%	\\
\hline
$M_{star}$ $\geq$ 10$^{10}$ 	& 82	& 68	& 98 \\
$M_{star}$ $\geq$ 10$^{9}$ 	& 73	& 53	& 90 \\
$M_{star}$ $\geq$ 10$^{8}$ 	& 56	& 38	& 80 \\
$M_{star}$ $\geq$ 10$^{7}$ 	& 48	& 30	& 70 \\
\hline
\noalign{\smallskip}
\hline
Differential			&	&	&	\\
\hline
$M_{star}$ range				& All	& E-S0a	& Sa-Im-BCD \\
M$_{\odot}$					& \%	& \%	& \%	\\
\hline
10$^{9}$ $\leq$ $M_{star}$ $\leq$ 10$^{10}$ 	& 68	& 42	& 89 \\
10$^{8}$ $\leq$ $M_{star}$ $\leq$ 10$^{9}$ 	& 41	& 27	& 65 \\
10$^{7}$ $\leq$ $M_{star}$ $\leq$ 10$^{8}$ 	& 20	& 6	& 38 \\
\noalign{\smallskip}
\hline
\end{tabular}
\]

}
\end{table}

The DS9/Funtools program \textit{Funcnts} has been run on all the extended UV detected sources catalogued in Voyer et al. (2014) (1771 objects). Counts have been transformed into
flux densities (in Jy) using the prescriptions given in Wright et al. (2010) and in the Explanatory Supplement \textit{WISE} Preliminary Data Release 
Products 
consistently with Ciesla et al. (2014), using 5.2269 $\times$ 10$^{-5}$ Jy/DN. We also applied, as suggested by Jarrett et al. (2013), an aperture correction of -0.03 mag
to account for the \textit{WISE} absolute photometric calibration method using PSF profile fitting. We also applied a second correction to account for a systematic difference 
in the calibrating "red" stars and "blue" galaxies. Jarrett et al. (2013) quantified this error for star-forming galaxies with a spectrum rising as $S(\nu)$ $\sim$ $\nu$$^{-2}$ 
and removed it by applying a systematic correction of 0.92 in the 22 $\mu$m band. We did not apply this correction to quiescent, early-type galaxies since in this band the 
emission might still be dominated by the Rayleigh-Jeans tails of the stellar atmosphere of M type stars. We did not apply any further colour correction 
since it is negligible in the $W4$ \textit{WISE} band ($\sim$ 1\%; Jarrett et al. 2013). Combined with calibration uncertainties ($\sim$ 1.5\%), the photometric uncertainty 
due to aperture and colour corrections on the $W4$ band 
should be of the order of$\sim$ 5\%. This, however, does not include the uncertainty on the total flux estimate, which in extended sources 
is generally dominated by the uncertainty on the determination of the sky background (Boselli et al. 2003b; Ciesla et al. 2012).
The total uncertainty on the measure of the flux density is thus given by the quadratic sum of the calibration uncertainty $err_{WISE}$, on the uncertainty of 
the large scale sky fluctuations $err_{sky}$ and on the pixel per pixel uncertainty $err_{pix}$ which might be partly correlated (Boselli et al. 2003b; Ciesla et al. 2012).
Consistently with Boselli et al. (2003b) and Voyer et al. (2014), we calculate the uncertainty on the sky $err_{sky}$ and on the pixel per pixel $err_{pix}$
in $\sim$ 10 square sky regions with randomly selected large sizes surrounding each target. These uncertainties are defined as:

\begin{equation}
{err_{sky} = N_{pix}STD[<b_1>...<b_n>]}
\end{equation}
 
\and 
\begin{equation}
{err_{pix} = \sqrt{N_{pix}}<STD[b_1]...STD[b_n]>}
\end{equation}
 
\noindent
where $N_{pix}$ is the number of pixels in the galaxy aperture, $STD[b_n]$ is the standard deviation of the values of all pixels in sky box $n$, and $<b_n>$ is the average 
of all pixels in sky box $n$ (Voyer et al. 2014). The total uncertainty on the extracted flux is then given by:

\begin{equation}
{err_{tot} = \sqrt{err_{sky}^2 + err_{pix}^2}}
\end{equation}

\noindent
to which the calibration uncertainty $err_{WISE}$ should be added for estimating the total photometric uncertainty on the data. Given that $err_{sky}$ and $err_{pix}$
are the dominant source of error, we consider here as detections all galaxies with $S_{22\mu m}$/$err_{tot}$ $>$ 1. 
As in Boselli et al. (2003b), for undetected galaxies we estimate an upper limit 
defined as:

\begin{equation}
{S_{22\mu m}(up. lim.) = 2 \times \sqrt{err_{sky}^2 + err_{pix}^2}}
\end{equation}

The \textit{WISE} 22$\mu$m flux densities of all the UV extended sources and their uncertainties are listed in Table \ref{TabWISE}, arranged as follow:

\begin{itemize}
\item {Column 1: Galaxy name, from NED.}
\item {Columns 2 and 3: Right ascension (J2000) and declination of the aperture used to extract the 22 $\mu$m \textit{WISE} flux density.}
\item {Columns 4, 5, and 6: Major and minor axis radii (in arcsec), and position angle (measured in degrees, from north clockwise) of the adopted aperture.}
\item {Column 7: flag indicating whether a galaxy is detected (1) or undetected (0).}
\item {Column 8: 22 $\mu$m \textit{WISE} flux density and error $err_{tot}$ as defined in eq. 9, in mJy. }
\end{itemize}

\begin{figure}
   \centering
   \includegraphics[width=9cm]{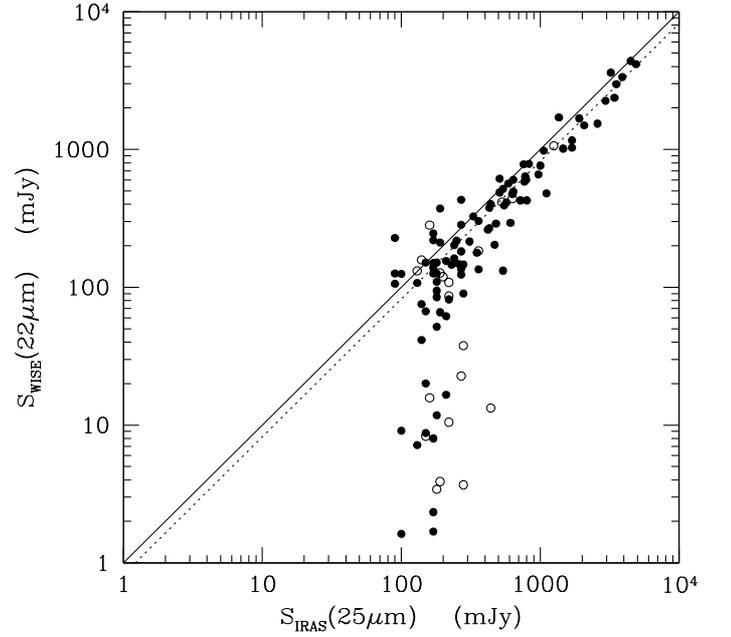}
   \caption{The comparison of the 22 $\mu$m \textit{WISE}
    flux densities determined in this work with those measured at 25 $\mu$m by 
   \textit{IRAS} for 119 detected galaxies in common. The solid line shows the 1:1 relation, while the dotted line the expected relation once the
   \textit{WISE} data are corrected by a factor of 1.22 as indicated by Ciesla et al. (2014) to take into account the shift in the photometric bands.
   Filled dots indicates late-type galaxies, empty-symbols early-types.}
   \label{IRAS}%
   \end{figure}

To check the quality of these \textit{WISE} data, we compare them to those already available in the literature in similar photometric bands.
We first use a compilation of 25 $\mu$m \textit{IRAS} data taken from different sources in the literature, available for 
119 galaxies of the sample (Boselli et al. 2010). The comparison of the two sets of data is shown in Fig. \ref{IRAS}. 
Figure \ref{IRAS} shows that for flux densities $S_{IRAS}(25 \mu m)$ $\gtrsim$ 400 mJy the \textit{WISE} and \textit{IRAS} sets of data 
are fairly consistent. Below this threshold, which roughly corresponds to the detection limit of \textit{IRAS} in the 25 $\mu$m band, 
the \textit{WISE} flux densities are systematically smaller than the \textit{IRAS} one, thereby suggesting that these \textit{IRAS} values are
probably spurious detections. By comparing 24 $\mu$m MIPS \textit{Spitzer} data to 22 $\mu$m \textit{WISE} data for the HRS,
Ciesla et al. (2014) found a systematic shift in the two sets of data of a factor 1.22 that they imputed to the 
slightly different spectral range covered by the two instruments. Figure \ref{IRAS} shows that the same systematic
shift can explain the observed difference in the \textit{IRAS} and \textit{WISE} data.

   \begin{figure}
   \centering
   \includegraphics[width=9cm]{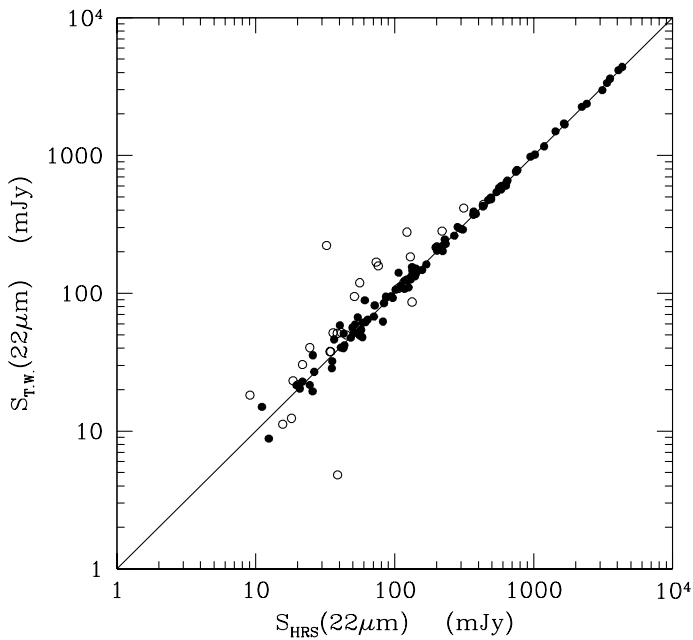}
   \caption{The comparison of the 22 $\mu$m \textit{WISE} flux densities for 138 HRS galaxies included in our sample, from Ciesla et al. (2014).
   Filled dots indicates late-type galaxies, empty-symbols early-types. The solid line shows the 1:1 relation.}
   \label{WISE}%
   \end{figure}

Figure \ref{WISE} shows the comparison of the 22 $\mu$m \textit{WISE} flux densities determined in this work with those recently
published by Ciesla et al. (2014) for the HRS galaxies in common (138 detected objects).
Figure \ref{WISE} shows that our set of data is perfectly consistent with the one of Ciesla et al. (2014). The mean Ciesla et al. (2014) to 
this work flux density ratio $S_{HRS}$/$S_{T.W.}$ for the 138 detected galaxies in common is $S_{HRS}$/$S_{T.W.}$ = 1.01 $\pm$ 0.63. 
The dispersion significantly drops when we limit the comparison to late-type systems ($S_{HRS}$/$S_{T.W.}$ = 0.99 $\pm$ 0.10; 105 objects), i.e. to
the galaxies where the \textit{WISE} data are crucial for an accurate dust extinction correction, while it is higher in early-type systems 
($S_{HRS}$/$S_{T.W.}$ = 1.06 $\pm$ 1.30; 33 objects). This systematic difference in early-type galaxies, where the 22 $\mu$m \textit{WISE} 
emission might still be dominated by the stellar emission and is thus limited to the innermost brightest regions, 
is due to the fact that while here we keep the same aperture than the one used to integrate 
the UV emission, in Ciesla et al. (2014) the apertures are manually adapted to encompass the total infrared emission on the \textit{WISE} images
and thus minimise the uncertainties due to the sky fluctuations. The procedure adopted in Ciesla et al. (2014) for early-type systems thus
should give more accurate results. We recall that the 22 $\mu$m flux densities of early-type galaxies 
are not used for the dust attenuation correction of the UV and optical photometric data. They are reported here 
just for completeness. 

Table \ref{WISEstat} gives the cumulative and differential detection rate in the 22 $\mu$m band in different bins of stellar mass for the whole sample 
of Virgo galaxies and separately for early- and late-type galaxies. The overall detection rate of late-type galaxies, where 22 $\mu$m flux densities are necessary for an accurate 
extinction correction, is fairly good ($\sim$ 70 \%) although it drops to 38 \% ~ in the lowest stellar mass bin. 

\end{appendix}

\begin{appendix}
\section{The stellar mass determination}

\begin{figure}
   \centering
   \includegraphics[width=9cm]{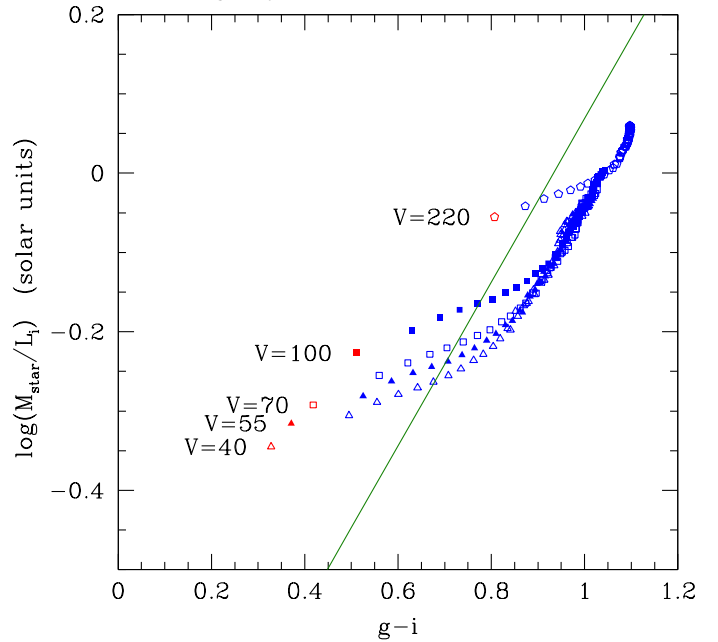}
   \caption{The relationship between the stellar mass-to-$i$-band luminosity ratio and the $g-i$ colour index as predicted by the calibration of Zibetti et al. (2009) (green
   solid line) and our models of galaxy evolution for unperturbed galaxies of different rotational velocity and fixed spin parameter ($\lambda$=0.05; red open symbols), as
   well as for galaxies undergoing a ram pressure stripping event (blue symbols). }
   \label{mstar2}%
   \end{figure}

The standard recipes such as those proposed by Bell \& de Jong (2001), Bell et al. (2003), Zibetti et al. (2009),
and Boselli et al. (2009) to estimate the stellar mass of galaxies using a combination of a stellar luminosity with an optical or near-infrared
colour index have been calibrated using different population synthesis models and IMF, and assuming different realistic star formation histories.
These star formation histories are generally assumed to reproduce the quite smooth evolution of unperturbed objects of different luminosity and morphological type. 
They are thus not ideally defined to reproduce the evolution of strongly perturbed galaxies in high-density environments such as those analysed in this work.
Indeed, in cluster galaxies the removal of the atomic and molecular gas content for a ram pressure stripping event is very rapid, and is thus able to
quench the activity of star formation on very short timescales. Thus, the standard recipes for determining the total stellar mass proposed in the literature
might not be optimised for perturbed galaxies such as those analysed in this work. Their adoption can induce systematic effects in the analysed sample.
To quantify these effects, we plot in Fig. \ref{mstar2} the relationship between the $i$-band mass-to-light ratio and the $g-i$ colour index for galaxies of
different stellar mass as predicted by our multizone chemo-spectrophotometric models of galaxy evolution and those predicted by the prescription of Zibetti et al. (2009).
Figure \ref{mstar2} shows a tight correlation between the two variables for unperturbed galaxies using either our evolutionary models (red symbols)
or the predictions of Zibetti et al. (2009), although this last gives a steeper relation. The observed difference in the two relations for unperturbed galaxies 
comes from the use of different population synthesis models, the adoption of different IMF and star formation histories of the target galaxies (e.g. 
Courteau et al. 2014). 
In this comparison, the main difference is that our model concerns disk galaxies while Zibetti et al. (2009)
is a fit to the M/L-color diagram obtained for a set of models including a very large variety of star formation
histories (not only adapted to star forming galaxies). The Zibetti et al. fit is thus an average of "active" galaxies (similar
to our bluest models, in the absence of interaction), and of "passive" galaxies (similar to our reddest models in which
the interaction has reduced the star formation activity).\\

\begin{figure}
   \centering
   \includegraphics[width=7cm]{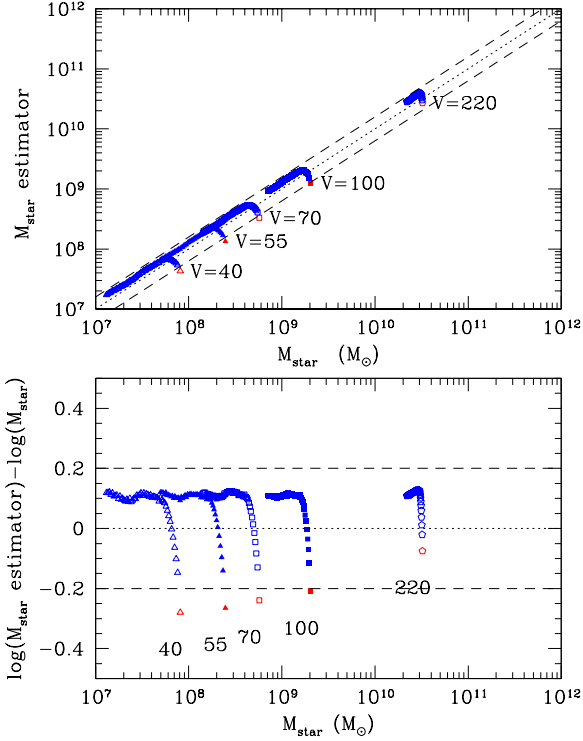}
   \caption{Upper panel: relationship between the stellar mass estimated using the Zibetti et al. (2009) mass-to-light colour dependent calibration and the real stellar mass
   of galaxies in our models of galaxy evolution for unperturbed (red symbols) and ram pressure stripped objects (blue symbols). The dotted line gives the 1:1 relation,
   while the dashed lines shows a variation of 0.2 dex (this value is typical of the global uncertainties affecting mass determinations due to e.g. the choice of the IMF).
   Lower panel: The logarithmic difference between the stellar mass estimated using the Zibetti et al. (2009) mass-to-light colour dependent calibration and the stellar mass
   from our models for unperturbed (red symbols) and ram pressure stripped objects (blue symbols) is plotted
   vs. the stellar mass of the model galaxies. }
   \label{mstar1}%
   \end{figure}

Figure \ref{mstar2} also shows that 
the relationship between $M_{star}/L(i)$ and $g-i$ significantly changes in perturbed galaxies. This is quite obvious since the colour of a galaxy significantly changes 
becoming redder once the galaxy has abruptly stopped its star formation activity, as indeed indicated by our models (Boselli et al. 2006; 2008a). If the $i$-band 
luminosity is barely affected after a ram pressure stripping event, the colour can significantly change on relatively short timescales. The adoption of a 
unique relation using standard recipes based on stellar luminosities and colours might thus induce strong systematic biases in the stellar mass determination.
It would be more appropriate to use a standard spectral energy distribution fitting code, provided that realistic truncated star formation histories such as those
observed in our sample can be easily reproduced. To effectively constrain the star formation history of galaxies, however, a full coverage of the UV-to-far-infrared spectral energy distribution
is necessary. This unfortunately is still quite prohibitive in the nearby universe for samples such as the one analysed in this work which spans a wide range in luminosity (from giant to dwarfs) 
and morphological type (from ellipticals to irregulars). Furthermore, the star formation history of the target galaxies, i.e. the topic of the present work, is an unknown variable, 
while not all fitting codes are tuned for such a purpose\footnote{These codes generally use parametrised star formation histories not always defined to reproduce the abrupt
truncation observed in cluster galaxies.}. We thus decided to estimate stellar masses using a standard recipe, and to quantify the uncertainty and any possible systematic
effects on the derived $M_{star}$ by comparing the prediction of our evolutionary models with the mass-to-light ratio vs. colour relations proposed in the literature.
This is done in Fig. \ref{mstar1}, where the $i$-band stellar mass-to-light ratio predicted by the Zibetti et al. (2009) prescription is plotted vs. the optical $g-i$ colour  
and compared to the predictions of our multizone chemo-spectrophotometric models.
 
\noindent
Figure \ref{mstar1} shows that, in unperturbed galaxies, the prescription of Zibetti et al. (2009) compared to our models 
underpredicts the stellar mass of galaxies by a factor of $\sim$ 0.3-0.1 dex
(larger values are for dwarf systems), 
while it overpredicts the mass in perturbed, gas deficient objects where the star formation activity is rapidly quenched after a 
ram pressure stripping event. In these objects, however, the overprediction is just by $\sim$ 0.1 dex irrespective of stellar mass. Figure \ref{mstar1} shows that,
although important, the systematic effect in the determination of the stellar mass using the recipe of Zibetti et al. (2009) is relatively small compared to the dynamic range 
in stellar mass sampled in this work. We thus decided to use the Zibetti et al. (2009) relation since it gives an "average" value for unperturbed and perturbed objects.
Figure \ref{mstar1} can be used to quantify the systematic effect on stellar mass for different galaxies.  

\end{appendix}

\onecolumn
\begin{center} \tiny \begin{longtable}{c c c c c c cc}
\caption{Example of \textit{WISE} data.\label{TabWISE}}\\
\hline \hline \\[-2ex]
\multicolumn{1}{c}{Name} &
\multicolumn{1}{c}{R.A.(J2000)$_{ap}$}&      
\multicolumn{1}{c}{dec$_{ap}$ }&    
\multicolumn{1}{c}{a} &     
\multicolumn{1}{c}{b} &    
\multicolumn{1}{c}{PA}  &
\multicolumn{1}{c}{flag} & 
\multicolumn{1}{c}{$S_{22 \mu m}$ } \\
\multicolumn{1}{c}{} &
\multicolumn{1}{c}{deg}&      
\multicolumn{1}{c}{deg} &    
\multicolumn{1}{c}{"} &     
\multicolumn{1}{c}{"} &    
\multicolumn{1}{c}{deg}  &
\multicolumn{1}{c}{} & 
\multicolumn{1}{c}{mJy}  \\
\multicolumn{1}{c}{(1)} &
\multicolumn{1}{c}{(2)}&      
\multicolumn{1}{c}{(3)} &    
\multicolumn{1}{c}{(4)} &     
\multicolumn{1}{c}{(5)} &    
\multicolumn{1}{c}{(6)}  & 
\multicolumn{1}{c}{(7)}  & 
\multicolumn{1}{c}{(8)}       \\[0.5ex] \hline
  \\[-1.8ex]
\endhead 
 \\[-1.8ex] \hline
  \multicolumn{7}{c}{{Continued on next page\ldots}} \\
\endfoot

  \\[-1.8ex] \hline \hline
\endlastfoot   
  CGCG-013019A    & 179.60542  &  0.70688    &  19.7     &   10.1    &   96      &  0 &        0.7		 \\ 
  CGCG-013019B    & 179.60883  &  0.71311    &  15.0     &   8.3     &  60       &  0 &        0.3		 \\ 
  CGCG-013019C    & 179.61542  &  0.71789    &  19.0     &   12.0    &   149     &  1 &        1.3  $\pm$     0.4\\ 
  CGCG-013034     & 180.07496  &  0.29497    &  30.8     &   12.6    &   102     &  1 &       22.3  $\pm$     0.6\\ 
  CGCG-013035     & 180.09963  &  0.49067    &  20.0     &   12.0    &   264     &  1 &        2.4  $\pm$     0.4\\ 
  CGCG-013036     & 180.27079  &  0.11375    &  25.0     &   10.0    &   225     &  1 &        1.6  $\pm$     0.5\\ 
  CGCG-013042     & 180.52258  &  1.52589    &  17.9     &   17.0    &   199     &  1 &        1.3  $\pm$     0.6\\ 
  CGCG-013045     & 180.67788  &  1.95219    &  28.1     &   13.8    &   231     &  1 &        2.8  $\pm$     0.4\\ 
  CGCG-013046     & 180.67462  &  1.97880    &  90.0     &   66.6    &   0       &  1 &      658.3  $\pm$     6.4\\ 
  CGCG-013049     & 180.80517  &  1.95092    &  25.0     &   20.0    &   151     &  1 &        2.3  $\pm$     0.7\\ 
  CGCG-013050     & 180.87304  &  1.22936    &  27.9     &   17.0    &   180     &  1 &        1.9  $\pm$     0.7\\ 
  CGCG-013051     & 180.90675  &  2.04692    &  20.0     &   13.0    &   187     &  0 &        0.8		 \\ 
  CGCG-013054     & 180.99783  &  1.41075    &  23.0     &   19.0    &   102     &  1 &       10.3  $\pm$     0.7\\ 
  CGCG-013055     & 181.02496  &  1.84697    &  13.0     &   30.0    &   9       &  1 &        1.1  $\pm$     0.7\\ 
  CGCG-013056     & 181.04058  &  1.82594    &  33.2     &   16.6    &   133     &  1 &        8.2  $\pm$     1.0\\ 
  CGCG-013057     & 181.08333  &  1.56758    &  70.8     &   15.6    &   189     &  1 &       29.1  $\pm$     1.1\\ 
  CGCG-013058     & 181.07917  &  1.84839    &  19.0     &   11.5    &   252     &  0 &        0.7		 \\ 
  CGCG-013059     & 181.11038  &  1.89660    &  56.7     &   43.9    &   20      &  1 &       11.7  $\pm$     2.2\\ 
  CGCG-013060     & 181.14192  &  1.73822    &  16.1     &   11.6    &   199     &  0 &        0.5		 \\ 
  CGCG-013061     & 181.14179  &  1.80161    &  35.0     &   15.0    &   249     &  0 &        1.1		 \\ 
  CGCG-013063     & 181.15829  &  1.78580    &  36.0     &   25.2    &   100     &  1 &       19.8  $\pm$     0.7\\ 
  CGCG-013064     & 181.15762  &  2.07253    &  45.0     &   23.1    &   216     &  1 &        7.0  $\pm$     1.3\\ 
  CGCG-013069     & 181.33979  &  2.08631    &  21.5     &   14.6    &   102     &  0 &        1.3		 \\ 
  CGCG-013072     & 181.41971  &  1.57506    &  37.2     &   10.4    &   187     &  0 &        1.3		 \\ 
  CGCG-013073     & 181.41942  &  1.59350    &  17.0     &   13.0    &   168     &  0 &        0.6		 \\ 
  CGCG-013076     & 181.63150  &  1.61656    &  20.0     &   18.0    &   120     &  1 &        6.8  $\pm$     0.5\\ 
  CGCG-013079     & 181.90867  &  1.57419    &  30.4     &   14.0    &   148     &  0 &        1.6		 \\ 
  CGCG-013080     & 181.99996  &  1.39594    &  25.0     &   20.0    &   210     &  1 &        2.8  $\pm$     0.6\\ 
  CGCG-013083     & 182.09800  &  0.11028    &  19.6     &   15.7    &   247     &  1 &        6.3  $\pm$     0.5\\ 
  CGCG-013084     & 182.13046  &  0.13650    &  28.2     &   15.8    &   200     &  1 &        8.9  $\pm$     0.8\\ 
  CGCG-013085     & 182.13021  &  1.90958    &  45.0     &   11.0    &   216     &  1 &        7.8  $\pm$     1.2\\ 
  CGCG-013087     & 182.36933  &  0.53325    &  19.2     &   10.0    &   169     &  1 &       15.6  $\pm$     0.4\\ 
  CGCG-013089     & 182.47446  &  0.92850    &  42.0     &   21.0    &   148     &  1 &       34.0  $\pm$     1.0\\ 
  CGCG-013095     & 182.63588  &  0.67750    &  37.9     &   19.7    &   161     &  0 &        2.3		 \\ 
  CGCG-013096     & 182.76692  &  0.97233    &  25.0     &   25.0    &   178     &  0 &        2.6		 \\ 
  CGCG-013100     & 182.99771  &  1.35003    &  40.0     &   17.0    &   258     &  1 &        3.7  $\pm$     1.5\\ 
  CGCG-013104     & 183.21725  &  1.30000    &  114.0    &    33.0   &    235    &  0 &       27.0		 \\ 
  CGCG-013105     & 183.42004  &  2.18769    &  22.0     &   19.0    &   230     &  0 &        4.3		 \\ 
  CGCG-013110     & 183.68938  &  0.74353    &  50.0     &   15.0    &   143     &  1 &       14.7  $\pm$     1.5\\ 
  CGCG-013113     & 183.96875  &  0.40069    &  50.0     &   37.0    &   244     &  1 &       22.5  $\pm$     4.6\\ 
  CGCG-013116     & 184.01867  &  1.18042    &  25.0     &   19.0    &   134     &  1 &       34.5  $\pm$     1.4\\ 
\end{longtable}
\end{center}

\twocolumn
\end{document}